\documentclass{emulateapj}
\usepackage{apjfonts}
\newcommand{\mbh}{M_{\rm BH}}
\newcommand{\mvir}{M_{\rm vir}}
\newcommand{\mbul}{M_{\rm sph}}
\newcommand{\nBH}{\dot{n}(\mbh)}
\newcommand{\nvir}{\dot{n}(\mvir)}
\newcommand{\nbul}{\dot{n}(\mbul)}
\newcommand{\nsigma}{\dot{n}(\sigma)}
\newcommand{\msigma}{\mbh-\sigma}
\newcommand{\zobs}{z_{\rm obs}}
\newcommand{\phistar}{\phi^{\ast}}
\newcommand{\Mstar}[1]{M^{\ast}_{#1}}

\newcommand{\etal}{et al.}

\newcommand{\Mdot}{\dot{M}}
\newcommand{\Lbol}{L_{\rm bol}}
\newcommand{\dEdt}{\epsilon_r \Mdot c^{2}}

\newcommand{\Lcut}[1]{10^{#1}\,L_{\sun}}

\newcommand{\dlgL}{{\rm d}\log(L)}

\newcommand{\Lp}{L_{\rm peak}}

\newcommand{\dtdL}{{\rm d}t/{\rm d}\log{L}}

\newcommand{\vvir}{V_{\rm vir}}

\newcommand{\fdphi}{\frac{{\rm d}\Phi}{{\rm d}\log L}}
\newcommand{\nLP}{\dot{n}(\Lp)}
\newcommand{\nLp}{\nLP}
\newcommand{\nstar}{\dot{n}_{\ast}}
\newcommand{\lstar}{L_{\ast}}
\newcommand{\sstar}{\sigma_{\ast}}

\newcommand{\lessim}{\lesssim}

\shorttitle{Red Galaxy Evolution from Quasars}
\shortauthors{Hopkins \etal}
\slugcomment{Accepted to ApJ, October, 2005}
\begin{document}

\title{Determining the Properties and Evolution of Red Galaxies
from the Quasar Luminosity Function}
\author{Philip F. Hopkins\altaffilmark{1}, 
Lars Hernquist\altaffilmark{1}, 
Thomas J. Cox\altaffilmark{1}, 
Brant Robertson\altaffilmark{1}, 
Volker Springel\altaffilmark{2}}
\altaffiltext{1}{Harvard-Smithsonian Center for Astrophysics, 
60 Garden Street, Cambridge, MA 02138, USA}
\altaffiltext{2}{Max-Planck-Institut f\"{u}r Astrophysik, 
Karl-Schwarzschild-Stra\ss e 1, 85740 Garching bei M\"{u}nchen, Germany}

\begin{abstract}

We examine the link between quasars and the red galaxy population
using a model for the self-regulated growth of supermassive black
holes in mergers involving gas-rich galaxies.  In this picture,
mergers drive nuclear inflows of gas, fueling starbursts and obscured
quasars until feedback energy from black hole growth expels the
surrounding gas, rendering the quasar briefly visible as a bright
optical source.  The quasar dies when there is no longer a significant
supply of gas to power accretion, and the stellar remnant relaxes as a
passively evolving spheroid satisfying the $M_{\rm BH}-\sigma$
relation and lying on the fundamental plane.  The same process that
halts black hole growth also terminates star formation in the remnant,
accounting for the observed red galaxy population in the bimodal
color/morphology distribution of galaxies.  Using a model for quasar
lifetimes and evolution motivated by hydrodynamical simulations of
galaxy mergers, we de-convolve the observed quasar luminosity function
at various redshifts to determine the birthrate of black holes of a
given final mass.  Identifying quasar activity with the formation of
spheroids in the framework of the merger hypothesis, this enables us
to infer the corresponding birthrate of spheroids with given
properties as a function of redshift.  With this method, we predict,
for the red galaxy population, the distribution of galaxy
velocity dispersions, the galaxy mass function, mass density, and star
formation rates, the luminosity function in many observed wavebands
(e.g., NUV, U, B, V, R, r, I, J, H, K), the total number density and
luminosity density of red galaxies, the distribution of colors as a
function of magnitude and velocity dispersion for several different
wavebands, the distribution of mass to light ratios as a function of
mass, the luminosity-size relations, and the typical ages and
distribution of ages (formation redshifts) as a function of both mass
and luminosity.  For each, we predict the evolution at redshifts 
$z=0-6$ and, in each case, our results are in good agreement with
observational estimates.  However, we demonstrate that the predictions
strongly disagree with observations if idealized, traditional models of
quasar lifetimes are adopted in which these objects turn on and off at
a fixed luminosity or follow simple exponential light curves, instead
of the more complicated quasar evolution implied by our simulations.

\end{abstract}


\section{Introduction}
\label{sec:intro}

Hierarchical theories of galaxy formation and evolution indicate that
large systems are built up over time through the merger of smaller
progenitors. Galaxy interactions in the local Universe motivate the
``merger hypothesis'' (Toomre \& Toomre 1972; Toomre 1977), according
to which collisions between spiral galaxies produce the massive
ellipticals observed at present times, a view supported by
self-consistent modeling of mergers (for reviews, see e.g.\ Barnes \&
Hernquist 1992; Barnes 1998).  Furthermore, it is believed that most
galaxies harbor supermassive black holes (e.g. Kormendy \& Richstone
1995; Richstone et al.\ 1998; Kormendy \& Gebhardt 2001) and that the
masses of these black holes correlate with either the mass (Magorrian
et al. 1998) or the velocity dispersion (i.e. the $M_{\rm
BH}$-$\sigma$ relation: Ferrarese \& Merritt 2000; Gebhardt et
al. 2000) of their host spheroids, demonstrating that the growth of
supermassive black holes and galaxy formation are linked. Simulations
of the self-regulated growth of black holes in galaxy mergers (Di
Matteo et al. 2005) have shown that the energy released by this
process can have a global impact on the structure of the remnant,
implying that models of galaxy formation and evolution must account
for black hole growth in a fully {\it self-consistent} manner.

Based on surveys such as SDSS, 2dFGRS, COMBO-17, and DEEP,
there is mounting evidence that the color distribution of galaxies at
$z=0$ is bimodal
\citep[e.g.][]{Strateva01,Blanton03,Kauffmann03a,Baldry04,Balogh04}, 
and can be well fitted by two Gaussians \citep[e.g.][]{Baldry04}. The
mean color and dispersion of these two (red and blue) distributions
depend on luminosity, but little on galaxy environment
\citep{Blanton03,Balogh04,Hogg04}. This bimodality extends to moderate
redshifts, $z\sim1.5$ 
\citep[e.g.,][]{Bell03,Bell04b,Willmer05,Faber05}, and there exists a
population of massive, very red galaxies at even higher redshift
\citep[e.g.][]{Franx03}.  The red galaxies in this bimodal
distribution are almost all elliptical, absorption-line galaxies, at
least at redshifts $z\lesssim1$
\citep[e.g.,][]{Strateva01,BernardiIV,Bell04a,Ball05}, which appear to be 
passively evolving from a redshift of peak star formation
$z\sim1.5-2.5$, according to both fundamental plane
\citep[e.g.,][]{vanDokkum01,Treu01,Treu02,Gebhardt03,Wuyts04,vandeVen03}, 
and color and spectral analyses
\citep[e.g.,][]{Menanteau01,Kuntschner02,Treu02,vandeVen03,Bell04b}. 
It also appears
that the properties of the red galaxies and their $z=0$
distribution, as well as their clustering and mass density evolution,
are consistent with their being formed through mergers and
thereafter relaxing quiescently
\citep[e.g.,][]{Kauffmann03b,Budavari03,Bell03,Baldry04,Weiner05}.

For mergers to produce red ellipticals from blue, star-forming disks
and yield a bimodal color distribution, the color must evolve rapidly,
or the observed bimodality would be washed out, requiring that star
formation be terminated soon after a merger.  \citet{SDH05a} showed
that this will not occur, especially in gas rich mergers at high
redshift, if black hole feedback is neglected, because even a small
amount of cold gas remaining after a powerful starburst will fuel a low
level of star formation for a Hubble time
\citep[e.g.][]{MH94, MH96, HM95}, preventing the remnant from
reddening sufficiently.  However, \citet{SDH05a} demonstrated that
feedback from black hole growth and quasar activity caused by mergers
can result in a much more violent and abrupt expulsion and heating of
the remaining gas, as the black hole nears its final mass.  This
process also produces a remnant that satisfies observed correlations
between black hole and host galaxy properties \citep{DSH05}.

Observations of elliptical galaxy ages and star formation histories
motivate the notion of ``anti-hierarchical'' growth, or ``cosmic
downsizing,''
\citep[e.g.,][]{Bower92,VF96,Ellis97,Bernardi98,Jorgensen96,Bell04b,Faber05},
where the most massive spheroids are also the oldest and reddest
systems.  While black hole feedback is likely a key ingredient in
shutting down star formation in these systems at high redshifts,
allowing them to redden onto the observed $z=0$ color-magnitude
relation, it does not automatically imply that particular black hole
and galaxy formation scenarios are self-consistent.  Moreover,
although there is evidence of downsizing in quasar activity, with the
most luminous quasars active at $z\sim2$ and the peak formation
redshift of quasars evolving as a function of luminosity
\citep[e.g.][]{Page97,Miyaji00,Cowie03, Ueda03,HMS05,LaFranca05}, it
has not been demonstrated that the implied downsizing is consistent or
even quantitatively similar to that of the galaxy population.  As we
demonstrate in what follows, the relationship between downsizing in
galaxy and quasar populations depends sensitively on the model chosen
for quasar light curves and lifetimes in any scenario in which
spheroids and quasars form together.

In our picture, red, remnant spheroids and supermassive black holes
are produced simultaneously in galaxy mergers which also yield starbursts
and quasar activity.  Previously, we studied black hole evolution in
mergers using simulations (Hopkins et al.\ 2005a-e), and showed that
the complex, luminosity-dependent quasar lifetimes and obscuration
\citep{H05a,H05b} lead to a new interpretation of the quasar
luminosity function \citep{H05c}, where the faint end of the
luminosity function consists mainly of quasars growing to much larger
final masses or in declining states following peak quasar activity.
This implies that the distribution of quasars being created at a given
redshift as a function of the quasar {\it peak} luminosity or {\it
final} black hole mass is {\em peaked} at a luminosity (mass)
corresponding to the observed {\em break} in the luminosity function,
falling off towards brighter and fainter luminosities. This differs
from all previous models of quasar lifetimes, which predict that this
distribution should have essentially identical shape to the observed
luminosity function, increasing monotonically with decreasing
luminosity (black hole mass).  Because our simulations also yield
observed correlations between black hole and remnant host galaxy
properties, we can deduce the distribution and evolution of the
remnant red galaxies produced in these merger events.  These
predictions will necessarily differ than those based on idealized
models of quasar lifetimes, which yield a qualitatively different
distribution of black hole masses (and thus host galaxy masses and
velocity distributions) being formed at any given redshift.

Here, we use our models of quasar lifetimes and lightcurves and the 
observed quasar luminosity function to
determine the rate at which quasars with a given peak luminosity or
final black hole mass are born in mergers.  Using the scaling
relations between black hole and host galaxy properties derived from
these simulations, we determine the birthrate of remnants with given
properties as a function of redshift, and use this to predict the
properties and evolution of the red, elliptical population in various
wavebands.  In \S~\ref{sec:methods} we describe our methodology,
including the simulations (\S~\ref{sec:sims}), our model of quasar
lifetimes and the quasar luminosity function (\S~\ref{sec:QLF}), and
the black hole-host galaxy scaling relations obtained from the
simulations (\S~\ref{sec:scaling}). In \S~\ref{sec:sigma} we use this
information to predict the distribution of galaxy velocity dispersions
with redshift, as well as the galaxy mass function and its evolution.
In \S~\ref{sec:gal.LFs} we obtain the galaxy luminosity function and
its evolution in many observed wavebands and for redshifts $z=0-6$. In
\S~\ref{sec:colors} we predict the distribution of galaxy colors as a
function of magnitude in several bands, velocity dispersion, and
redshift. In \S~\ref{sec:ML} we estimate the distribution of
mass-to-light ratios and luminosity-size relation, and their
differential evolution with time, as a function of mass and
redshift. In \S~\ref{sec:ages} we predict the distribution of
formation ages (redshifts) as a function of galaxy mass, velocity
dispersion, and luminosity.  Finally in \S~\ref{sec:conclusions} we
discuss our results and their implications for observations and models
of the joint formation of spheroids and active galactic nuclei (AGN).

Throughout, we adopt a $\Omega_{\rm M}=0.3$, $\Omega_{\Lambda}=0.7$,
$H_{0}=70\,{\rm km\,s^{-1}\,Mpc^{-1}}$ cosmology. Unless otherwise
stated, all magnitudes are in the Vega system.

\section{Methodology}
\label{sec:methods}
\subsection{The Simulations}
\label{sec:sims}

The simulations were performed using {\small GADGET-2}
\citep{Springel2005}, a new version of the parallel TreeSPH code
{\small GADGET} \citep{SYW01}.  {\small GADGET-2} employs a fully
conservative formulation \citep{SH02} of smoothed particle
hydrodynamics (SPH), which maintains simultaneous energy and entropy
conservation even when smoothing lengths evolve (see e.g., Hernquist
1993b, O'Shea et al. 2005). Our simulations account for radiative
cooling, heating by a UV background (as in Katz et al. 1996, Dav\'e et
al. 1999), and incorporate a sub-resolution model of a multiphase
interstellar medium (ISM) to describe star formation and supernova
feedback \citep{SH03a}. Feedback from supernovae is captured in this
sub-resolution model through an effective equation of state for
star-forming gas, enabling us to evolve disks with
large gas fractions so that they are stable
against fragmentation (see, e.g.\ Springel et al.\ 2005b; Springel \& Hernquist
2005; Robertson et al.\ 2004, 2005a).

Supermassive black holes (BHs) are represented by ``sink'' particles
that accrete gas at a rate $\Mdot$ estimated from the local gas
density and sound speed using an Eddington-limited prescription based
on Bondi-Hoyle-Lyttleton accretion theory (Bondi 1952; Bondi \& Hoyle
1944; Hoyle \& Lyttleton 1939).  The bolometric luminosity of the
black hole is $\Lbol=\dEdt$, where $\epsilon_r=0.1$ is the radiative
efficiency.  We assume that a small fraction (typical $\approx 5\%$)
of $\Lbol$ couples dynamically to the surrounding gas, and that this
feedback is injected into the gas as thermal energy.  This fraction is
a free parameter, which we determine as in \citet{DSH05} by matching
the observed normalization of the $M_{\rm BH}-\sigma$ relation.  For
now, we do not resolve the small-scale dynamics of the gas directly
around the black hole, but assume that the time-averaged accretion
rate can be estimated on the scale of our spatial resolution (reaching
$\approx 20$\,pc, in the best cases).

The progenitor galaxies are constructed as described in
\citet{SDH05b}.  For each simulation, we generate two stable, isolated
spiral galaxies, with dark matter halos having a \citet{Hernquist90}
profile, motivated by cosmological simulations (e.g. Navarro et
al. 1996; Busha et al. 2004), simple analytical arguments (e.g. Jaffe
1987; White 1987; see Barnes 1998, \S 7.3), and observations
(e.g. Rines et al. 2002, 2002, 2003, 2004), an exponential disk of gas
and stars, and (optionally) a bulge.  The galaxies have masses $M_{\rm
vir}=V_{\rm vir}^{3}/(10GH_{0})$ for $z=0$, with the baryonic disk
having a mass fraction $m_{\rm d}=0.041$, the bulge (when present) has
$m_{\rm b}=0.0136$, and the rest of the mass is in dark matter
typically with a concentration parameter $9.0$.  The disk scale-length
is computed based on an assumed spin parameter $\lambda=0.033$, chosen
to be near the mode in the observed $\lambda$ distribution
\citep{Vitvitska02}, and the scale-length of the bulge is set to $0.2$
times the resulting value.

Typically, each galaxy is initially composed of 168000 dark matter
halo particles, 8000 bulge particles (when present), 24000 gas and
24000 stellar disk particles, and one BH particle. We vary the
numerical resolution, with many of our simulations using instead twice
as many particles in each galaxy, and a subset of simulations with up
to 128 times as many particles.  We vary the initial seed mass of
the black hole to identify any systematic dependence of our results on
this choice.  In most cases, we choose the seed mass either in accord
with the observed $M_{\rm BH}$-$\sigma$ relation or to be sufficiently
small that its presence will not have an immediate effect.  Given the
particle numbers employed, the dark matter, gas, and star particles
are all of roughly equal mass, and central cusps in the dark matter
and bulge profiles are reasonably well resolved (see Fig 2. in
Springel et al. 2005b).

The form of our fitted quasar lifetimes and galaxy scaling relations
are based on a series of several hundred merger simulations, described
in \citet{Robertson05b} and \citet{H05e}.  We vary the resolution, the
orbital geometry, the masses and structural properties of the merging
galaxies, the mass ratio of the galaxies, initial gas fractions, halo
concentrations, the parameters describing star formation and feedback
from supernovae and black hole growth, and initial black hole masses.
The progenitors have virial velocities $\vvir=80, 113, 160, 226, 320,
{\rm and}\ 500\,{\rm km\,s^{-1}}$, constructed to resemble galaxies at
redshifts $z=0, 2, 3, {\rm and}\ 6$, and span a range in final black
hole mass $\mbh\sim10^{5}-10^{10}\,M_{\sun}$.  This large set of runs
allows us to investigate merger evolution for a wide range of galaxy
properties and to identify any systematic dependence of our modeling.
Moreover, the extensive range of conditions probed gives us a large
dynamic range in our simulations, with final spheroid masses spanning
$\mbul\sim10^{8}-10^{13}\,M_{\sun}$, covering nearly the entire observed
range.

\subsection{Quasar Lifetimes and the Quasar Luminosity Function}
\label{sec:QLF}
\nobreak
Previous theoretical studies of the quasar luminosity function have
generally employed idealized quasar light curves, either some variant
of a ``feast or famine'' or ``light bulb'' model 
\citep[in which quasars
have only two states: ``on'' or ``off'', with constant luminosity in
the ``on'' state; e.g.,][]{SB92,KH00,HM00,HQB04} or a pure
exponential light curve 
\citep[constant Eddington-ratio growth or
exponential decay; e.g.,][]{HL98,V03,WL03}.  However, our
simulations of galaxy mergers suggest that these models are a poor
approximation to the quasar lifetime at any given luminosity.  The
light curves from the simulations are complex, generally having
periods of rapid accretion after ``first passage'' of the galaxies,
followed by an extended quiescent period, then a transition to a peak,
highly luminous quasar phase, and then a dimming as self-regulated
mechanisms expel gas from the remnant center after the black hole
reaches a critical mass. In addition, the accretion rate at any time
can be variable over small timescales $\sim{\rm Myr}$, but despite
these complexities, the statistical nature of the light curve can be
described by simple forms, which we describe below.

From the simulations, we find that the differential quasar lifetime,
i.e.\ the time spent by a quasar in a merger in a given logarithmic
luminosity interval, is well fitted by an exponential,
\begin{equation}
\dtdL=t^{\ast}_{Q}\, \exp [-L/L^{\ast}_{Q}], 
\end{equation}
where $L^{\ast}_{Q}$ is proportional to the {\em peak} quasar
luminosity ($\Lp$; roughly, the Eddington luminosity of the final
black hole mass), and $t^{\ast}_{Q}$ is weakly dependent on peak
luminosity.  When quantified as a function of $\Lp$ in
this manner, the quasar lifetime shows no systematic dependence on any
host galaxy properties, merger parameters, initial black hole masses,
ISM and gas equations of state and star formation models, or any other
varied parameters (Hopkins et al. 2005e).

If quasars of a given peak luminosity are being created or activated
at a rate $\nLp$ at some redshift $z$, then, to first order in the
quasar lifetime over the Hubble time, the observed quasar luminosity
function (neglecting attenuation) is
\begin{equation}
\phi(L)\equiv\fdphi(L)=\int{\frac{{\rm d}t(L,L_{\rm
      peak})}{\dlgL}\,\nLP}\,{\rm d}\log(L_{\rm peak}). 
\end{equation}
Knowing the quasar lifetime, we can invert this relation to
determine the birthrate of quasars as a function of peak
luminosity and redshift, $\nLp$.  As shown in
\citet{H05e}, the quasar luminosity functions in optical, UV, 
soft X-ray, and hard X-ray 
wavebands (including the effects of extinction) and at all measured
redshifts are simultaneously well-fitted by a lognormal $\nLp$,
\begin{equation}
\nLP=\nstar\ \frac{1}{\sstar\sqrt{2\pi}}\exp\Bigl[ -\frac{1}{2}\,\Bigl( \frac{\log(\Lp/\lstar)}{\sstar}\Bigr)^{2}\Bigr].
\label{eqn:nLp.lognorm}
\end{equation}
Here, $\nstar$ is the total number of quasars being created
or activated per unit comoving volume per unit time; $\lstar$ is the
median of the lognormal, the characteristic peak luminosity of quasars
activating (i.e.\ the peak luminosity at which $\nLP$ itself peaks),
which is directly related to the break luminosity in the observed
luminosity function \citep{H05c}; and $\sstar$ is the width of the
lognormal in $\nLp$, which determines the slope of the bright end of the
luminosity function.  The evolution of the quasar luminosity function
with redshift is well-described by pure peak-luminosity evolution,
where $\nstar$ and $\sstar$ are constant but
$\lstar=\lstar^{0}\,\exp({k_{1}\,\tau})$. 
Here, $\tau$ is the
fractional lookback time $\tau=H_{0}\,\int^{0}_{z}dt$.  
Above $z\sim2-3$, the 
quasar population declines, but the detailed shape and evolution of the 
faint-end of the quasar luminosity function at these redshifts is poorly 
constrained from observations.
Therefore we consider two choices: either 
pure peak-luminosity evolution (PPLE), where we multiply 
$\lstar$ by a
factor $\exp({-k_{2}\,[z-2]})$ for $z>2$, or pure density evolution (PDE), where 
we multiply the $z=2$ luminosity function by a normalization factor 
(i.e.\ multiply $\nstar$ by a factor) with identical functional form.

We follow \citet{H05e}, but fit to the more recent luminosity
functions in the hard X-ray, soft X-ray, and optical from
\citet{Ueda03,HMS05,Richards05}, respectively, and find the best-fit
parameters $(\log(\lstar^{0}/L_{\sun}),\,k_{1},\,k_{2},\,\log(\nstar/\
{\rm Mpc^{-3}\,Myr^{-1}}),\,\sstar)=
(11.3,\,4.0,\,0.65,\,-6.37,\,0.7)$.  These are similar to the values
given by \citet{H05e} using older observations, although they suggest
a somewhat narrower width in peak luminosities (with the peak more
closely related to the break in the observed luminosity function).

From the form of the quasar light curve and lifetime as a function of
luminosity, we can calculate the final black hole mass for a given
$\Lp$, and convert from $\nLp$ to $\nBH$, the birthrate of
black holes of a given final (post-merger) mass.  Accounting for the
corrections owing to the non-trivial shape of the quasar light curve and
lifetime, we obtain
\begin{equation}
\mbh= M_{\rm Edd}(\Lp)\ [1.24\, (\Lp/\Lcut{13})^{-0.11}].
\end{equation}
Applying this conversion to our fitted $\nLp$, we find that $\nBH$ is
also a lognormal, with identical redshift evolution and functional
form, and $(\log (M_{\ast,\rm BH}^{0}/M_{\sun}),\, k_{1},\, k_{2},\,
\log(\dot{n}_{\ast,\rm BH}/\ {\rm
Mpc^{-3}\,Myr^{-1}}),\,\sigma_{\ast,\rm BH})=
(6.45,\,3.2,\,0.59,\,-6.25,\,0.62)$.

\begin{figure}
    \centering
    %\plotone{show.LF.ps}
    \epsscale{1.15}
    \plotone{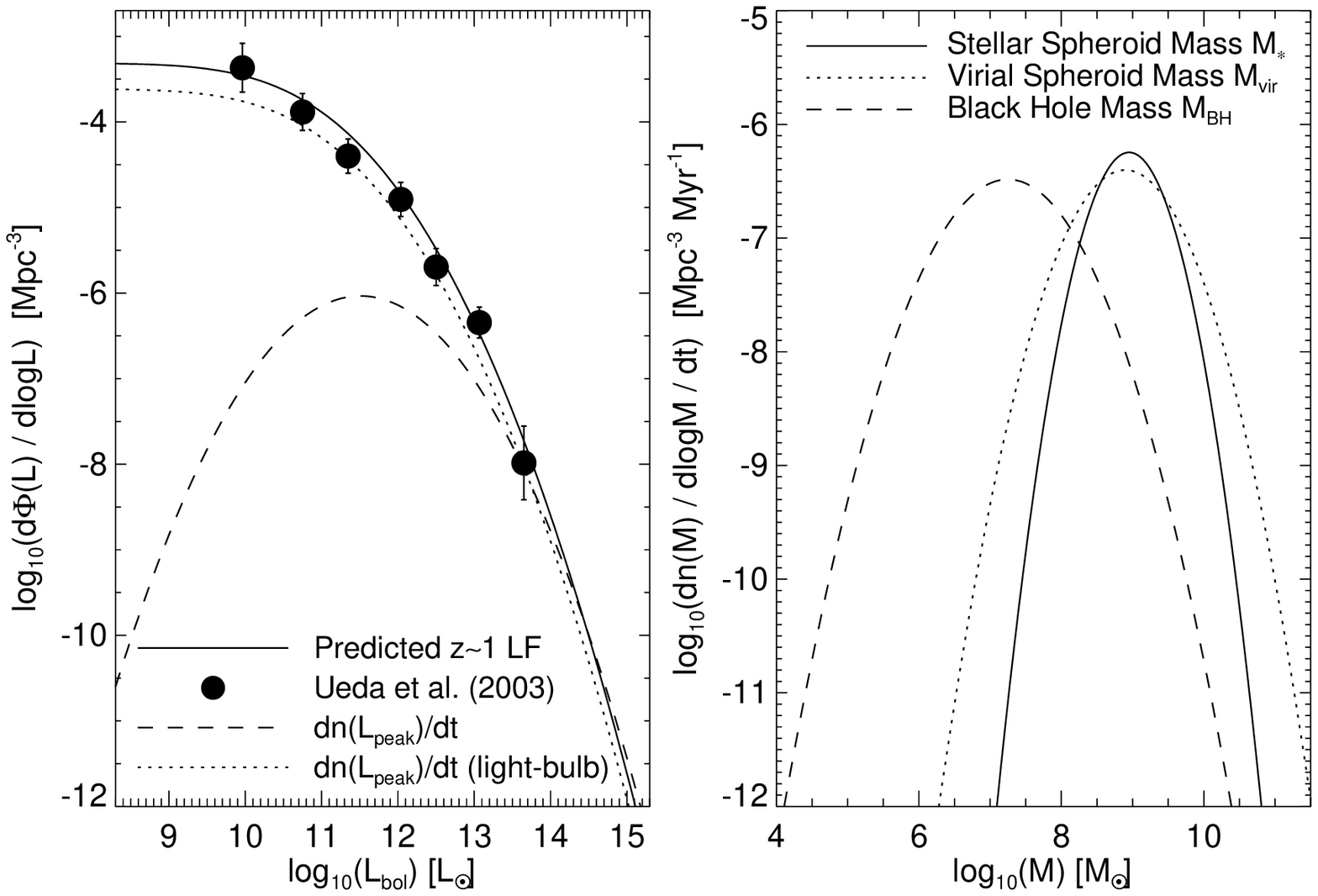}
   \caption{Predicted luminosity function (left, solid line) at $z=1$
   using our model for quasar lifetimes and evolution.
   The corresponding hard X-ray luminosity function of
   \citet{Ueda03} (circles) is shown for comparison, rescaled to
   bolometric luminosity following \citet{H05e,Marconi04,VBS03}. The
   $\nLp$ distribution (rescaled in arbitrary units for comparison) is
   shown (dashed line), as is the $\nLp$ distribution obtained using a
   ``light-bulb'' or exponential light curve model of the
   quasar lifetime (dotted). On the right, the corresponding rate of
   formation of black holes/quasars of a given final mass, $\nBH$ is
   shown (dashed), as well as the rate of formation of remnant
   spheroids of a given virial (dotted; $\nvir$) and stellar (solid;
   $\nbul$) mass, determined from the $\mbh-\mvir$ and fundamental
   plane relations of our simulations (Robertson et al. 2005b).
    \label{fig:show.LF}}
\end{figure}

Figure~\ref{fig:show.LF} shows an example of the results of our
procedure for deconvolving an observed quasar luminosity function to
obtain the black hole birthrate, using the \citet{Ueda03} hard X-ray
luminosity function at $z\sim1$. The left panel gives the quasar
luminosity function, where the black points are the observations and
the line is the prediction from the quasar lifetimes and fitted $\nBH$
above. The right panel shows the corresponding $\nBH$ distribution at
this redshift.  The $\nLp$ [$\nBH$] distribution derived has the
property that it {\em peaks} at a characteristic peak luminosity
(black hole mass) corresponding to the break in the observed
luminosity function, and falls off to both lower and higher
luminosities.  In this interpretation of the quasar luminosity
function, ``cosmic downsizing'' follows naturally as the break in the quasar
luminosity function moves to lower luminosities at lower redshifts,
and the implied downsizing is indeed quantitatively more dramatic than
that implied by idealized models of quasar activity (see \S~\ref{sec:ML}
and Figure 23 of Hopkins et al.\ 2005e).

It is important to note that the slope of the faint (low-$\mbh$) end
of $\nBH$ is only weakly constrained by the observed luminosity
function, a point discussed further in \S~\ref{sec:LF.B}.  To
illustrate this, the figure shows the birthrate of quasars of a given
peak luminosity, $\nLp$ (plotted in arbitrary units to demonstrate
this qualitative behavior) as the dashed line.  The $\nLp$
distribution which would be obtained using a ``light-bulb'' or
exponential light curve model of the quasar lifetime is also shown
(dotted line) for comparison (the $\nBH$, $\nvir$, $\nbul$
distributions for such a model will have the same shape as the $\nLp$
distribution, as explained below in \S~\ref{sec:scaling}).  The two
models make very different predictions for luminosities/masses below
those corresponding to the break in the observed luminosity function.

Although we do not consider the brief active quasar and starburst
phases in our subsequent analysis (as they are heavily affected by
rapidly evolving star formation, dust obscuration, merger dynamics,
and quasar luminosities), we note that our modeling allows us to
predict the behavior of the active quasar host galaxy luminosity
function.  We expect that the active quasars at a given redshift
should have a narrow range in peak luminosities (black hole masses),
corresponding to a narrow range in host galaxy stellar masses. This is
shown in the $\nLp$ and $\nbul$ (derived below) distributions given in
Figure~\ref{fig:show.LF}. We therefore expect that the host galaxies
of quasars active at a given time will have a much narrower range in
luminosities than that predicted by e.g.\ idealized models of the quasar
lifetime (for which $\nLp$ and therefore $\nbul$ must increase
monotonically with decreasing luminosity/mass; see e.g. Lidz et al.
2005). There is observational support for this: the quasar host galaxy
luminosity function is found to follow an approximately lognormal
distribution with narrow width $\sigma_{\log(L,\,{\rm host})}\sim
\sigma_{\log(M,\,{\rm host})}=0.2$ ($\sim0.6-0.7$ magnitudes) and a
peak roughly corresponding to the stellar mass of quasar hosts with
$L_{peak} \sim $ the quasar luminosity function break luminosity
\citep{Bahcall97,McLure99,Hamilton02}.

\subsection{Scaling Relations Among Galaxy and Black Hole Properties}
\label{sec:scaling}

Self-regulated black hole growth in our simulations 
yields a black hole mass-bulge velocity
dispersion ($\msigma$) relation \citep{DSH05} which agrees well
with the observations of, e.g.\ \citet{Gebhardt00,FM00,Tremaine02}.
\citet{Robertson05b} further study this relation,
and find it holds for mergers occurring at 
any redshift, 
with a constant slope and weak evolution in the normalization. 
From the simulations, they find
\begin{equation}
\log{(\mbh/M_{\sun})}\approx8.1+4.0\,\log\Bigl( \frac{\sigma}
{200\,{\rm km\,s^{-1}}}\Bigr)-0.19\,\log(1+z) .
\end{equation}
The precise values depend on the fitting
method, but in all cases agree well with those determined in
\citet{Tremaine02} for $z=0$. The scatter about this relation from the
simulations is $\sim0.3$\,dex, similar to that observed. 
It is also important to note that the evolution seen in
these simulations produces a $z=0$ scatter consistent with what is
observed, which is not the case for all theoretical models
\citep{Robertson05b}.

The weak evolution in
the $\msigma$ relation is caused by an increasing $\sigma$ for a given
stellar mass with increasing redshift, as halos at higher redshift are
more compact; 
the relation between black hole mass and total stellar mass 
($\mbh-\mbul$) is independent of
redshift.  This independence is also suggested observationally by
galaxy-AGN clustering properties as a function of redshift
\citep{AS05}.  From our simulations, we can
similarly determine the $\mbh-\mvir$ and $\mbh-\mbul$ 
relationship, giving
\begin{equation}
\mbh=7.0\times10^{-4}\,\mvir, 
\end{equation}
\begin{equation}
\mbh=0.001\,\mbul,
\end{equation}
in reasonable agreement with the 
observations of \citet{MarconiHunt03}, if we account for the 
slightly different definitions of $\mvir$ used.
Here, $\mvir$ is the virial mass within an effective radius, 
alternatively defined by $\mvir=k\,\sigma^{2}R_{e}/G$, where to be definite we 
take $\sigma$ to be the average spheroid velocity dispersion within the effective 
radius $R_{e}$. For this conversion (where necessary) we adopt $k=5$, as 
is roughly seen in our simulations and expected 
for e.g.\ a Hernquist (1990) spheroid or $R^{1/4}$-law profile, 
and also similar to that 
suggested by comparison of mass measurements from dynamical modeling 
and from measurements of $\sigma$ and $R_{e}$ (e.g.\ Cappellari et al.\ 2005, 
although compare Marconi \& Hunt 2003, 
who adopt $k=3$, which is the primary reason for the 
small discrepancy in the relation they observe and those we show above). 
The scatter about this relation from the simulations is
small, about $\sim0.3$\,dex, similar to that observed.

We note that there are considerable observational contradictions 
regarding possible evolution in the $M_{\rm BH}-\sigma$ or 
$M_{\rm BH}-\mbul$ relations, with e.g.\ \citet{Shields03} and 
\citet{AS05} finding no evolution to $z\sim3$ and 
e.g.\ \citet{Peng05} and \citet{McLure05} finding 
substantial evolution at $z<2$ (specifically, substantially 
undermassive bulges at $z\sim2$). However, these observations are 
still difficult and have large uncertainties; furthermore, they 
specifically select primarily active, high Eddington ratio objects, 
which local observations \citep[e.g.,][]{Barth05} suggest may be biased 
to lie above the $M_{\rm BH}-\sigma$ relation in the manner observed. 
Above $z\sim2$, the possibility for such evolution, and 
the uncertainty resulting from it, is essentially captured in our consideration 
of pure luminosity vs.\ pure density evolution for the quasar luminosity function, since 
these different evolutions imply a different peak luminosity 
(i.e.\ final spheroid mass) distribution. Thus, the uncertainties introduced 
by such evolution are not significantly larger than those we already 
describe, unless there is large evolution at $0<z<2$. Even such evolution 
in the $M_{\rm BH}-\mbul$ relation will not change many of our conclusions, 
if the stellar mass of the final spheroid is primarily 
{\em formed} at this time, but is simply assembled (presumably in 
subsequent dry mergers) at later times. The alternative, that 
this stellar mass is formed between $z=2$ (when the massive black holes
were formed) and $z=0$, is ruled out strongly by many observations 
which show the host spheroids of these black holes have old stellar 
populations with redshifts of formation $z\sim1.5-2.5$ 
\citep[e.g.][]{Bower92,Jorgensen96,VF96,Ellis97,
Bernardi98,Jorgensen99b,vandeVen03,Cross04,Wuyts04,Bell04b,FS04,Labbe05}.

Robertson et al.\ (2005c, in preparation) employ our simulations
to study the
fundamental plane relation between spheroid effective radius $R_{e}$,
velocity dispersion $\sigma$, and stellar surface mass density $\Sigma$
of the merger remnants.   In this,
the projected stellar surface density $\Sigma$ is calculated
along many different lines-of-sight, and for each, $R_{e}$ is determined
as the two-dimensional radius enclosing half the 
stellar mass, and $\sigma$ is the mass-weighted 
line-of-sight stellar velocity dispersion within an aperture of radius $R_{e}$.
When compared to e.g.\ the observed $K$-band fundamental place, 
for which a constant mass-to-light ratio is a reasonable approximation, 
the remnant spheroids of gas-rich mergers from our simulations fall on the
observed infrared 
fundamental plane \citep[$R_{e}\propto\sigma^{1.53}\Sigma^{-0.79}$,
e.g.,][]{Pahre98a,Pahre98b} with little
scatter. This relation and direct measurement yields a 
stellar mass-effective radius relation of the form 
$R_{e}\propto\mbul(R_{e})^{\beta}$ (where $\mbul(R_{e})$ is the stellar mass within 
the effective radius), or 
\begin{equation}
\log(R_{e}/{\rm kpc})=\alpha+\beta\,\log(\mbul(R_{e})/M_{\sun}). 
\end{equation}
The average $\mbul(R_{e})-R_{e}$ relation found in our simulations has best-fit coefficients 
$\alpha=-5.81$,\ $\beta=0.57$, (i.e.\ 
$R_{e}\approx0.86\,{\rm kpc}\,(\mbul(R_{e})/10^{10}\,M_{\sun})^{0.57}$), 
in good agreement with observations
\citep{Shen03} after accounting for the 
small difference between effective radius used here and half-light radius observed. 
The exact relation has a weak dependence on redshift, which we 
include; but we find little difference in our results in either case as, for example, at 
$z=0$, $\alpha=-5.6,\ \beta=0.56$, and at $z=2$, 
$\alpha=-5.4,\ \beta=0.53$. Observations also suggest only
weak evolution in this relation
\citep[e.g.][]{Trujillo04a,Trujillo04b,Trujillo05,McIntosh05b}. 

We use this relation to convert between mass-to-light ratios as a function of stellar mass to 
a luminosity-size relation in \S~\ref{sec:ML}, but we can also use it 
to determine the spheroid stellar mass as a function of virial (dynamical) mass 
and black hole mass. Combining the equations above,
\begin{equation}
\frac{\mbul(R_{e})}{\mvir}
\approx0.3\,\Bigl( \frac{\mvir}{10^{10}\,M_{\sun}} \Bigr)^{-0.2}. 
\end{equation}
This agrees well with observations (e.g. Bernardi et al.\ 2003a, Padmanabhan
et al.  2004; Cappellari et al. 2005) and additionally follows from the observed 
$\mbul-R_{e}$ relation given that $M_{\rm BH}\propto \sigma^{4}\propto \mvir$.
Note that we have defined the stellar mass $\mbul(R_{e})$ as that within the effective radius 
$R_{e}$; this means that 
the {\em total} galaxy stellar mass is $\mbul\approx2\,\mbul(R_{e})$.
These relations are determined
from the simulations to be independent of redshift (except for the weak 
evolution in $\mbul-R_{e}$ which we account for).
When only the total 
stellar mass is needed, we use the directly fitted $\mbh-\mbul$ relation 
described above as it both avoids the uncertainties inherent in these conversions and 
accounts for e.g.\ changing bulge-to-disk ratios as a function of mass. 

In what follows, we are not concerned with the structure of
individual galaxies, and so defer a detailed structural analysis of
merger remnants (Robertson et al.\ 2005c, in preparation).  We instead
use the relations above to study the statistical properties (i.e.\
conditional age and mass distributions) and evolution of the red
galaxy population.  We emphasize that although we use the form of
these relations from our simulations, because each agrees well with
its observed counterpart, it makes no difference to our results
whether we use the relations from our simulations or adopt the
observed scalings.

The simulations yield relationships between black hole mass and either
velocity dispersion or stellar spheroid mass that can be used to
transform the birthrate of black holes of a given final mass $\mbh$,
$\nBH$ into a birthrate of remnants with definite velocity dispersion
$\nsigma$ or stellar spheroid mass $\nbul$. This is illustrated in
Figure~\ref{fig:show.LF}, where the right panel gives the $\nvir$
(dotted) and $\nbul$ (solid) relations derived using the fitted
relations above, our modeling of quasar lifetimes, and the observed
quasar luminosity function. Although there are several steps in this
procedure, we emphasize that all of the relationships used, each
agreeing with observations, are determined entirely from the
simulations alone, in a self-consistent manner.  Any additional
modeling required beyond this point is further calculated
self-consistently from the simulations and is directly constrained by
observations of quasars (e.g.\ the cases of obscuration and quasar
lifetimes; Hopkins et al.\ 2005e) or galaxies (e.g.\ star formation
and stellar population synthesis models; Bruzual \& Charlot 2003).
The lone observational input is the observed {\em quasar} luminosity
function, from which we derive the birthrate of spheroids of a given
mass (or velocity dispersion).

In our simulations, merger remnants resemble elliptical galaxies, with
small gas fractions, and star formation is terminated by
feedback as the black hole reaches its final mass.  Thereafter, the
galaxies mainly evolve passively, without significant star formation.
The timescale for the merger-induced starburst is $\sim100$\,Myr
(e.g. Springel et al. 2005a), much shorter than the merger timescale
$\sim$Gyr.  We therefore adopt the approximation that the merger
occurs instantly at the redshift being considered, and that the
remnant does not evolve after that point (at least to very high
redshifts where the Hubble time becomes comparable to the timescale
for the merger).  We have actually considered two cases: one
where we assume each spheroid is formed instantly at the redshift
under consideration, and a second where we assume the starburst to
have a Gaussian shape in time with a peak at $z$ and characteristic
falloff timescale (standard deviation) $\sim100$\,Myr.  We find
essentially no difference in our predictions between these cases,
except for a slight reddening of typical galaxy colors at high
redshift in the latter case.  We also do explicitly calculate the
possible consequences of subsequent ``dry merging'' in
\S~\ref{sec:LF.B} below, and show that they are small.

Given the birthrate of spheroids, we use the stellar population
synthesis models of \citet{BC03} to determine their observed
luminosities and colors.  The remnants in our simulations typically
have solar metallicities, even at high redshift, \citep[as expected
from observations of high-redshift red galaxies,
e.g.][]{vanDokkum04,FS04} as metal enrichment occurs through star
formation and associated supernova feedback in the most dense regions
of the galaxy and metals are distributed throughout the galaxy by
quasar feedback \citep{Cox05}.  To examine the impact of the
metallicity on the stellar population, we consider two cases:
one in which the remnants are assumed to have solar metallicity
($0.02$) and the second where they have a Gaussian metallicity
distribution (with a mean solar metallicity) and standard deviation
$\sim0.005-0.01$.  We find little difference in our results between
these two cases.

A scaling of metallicity with mass or velocity dispersion $\sigma$
could also influence our predicted luminosity functions.  There is
some observational evidence of a correlation between metallicity and
$\sigma$ \citep[e.g.][] {Worthey92,Jorgensen97,Kuntschner00}, but the
inferred metallicities are degenerate with the modeled population ages
\citep{Worthey95,Faber95,Worthey97} and some studies infer no
connection between metallicity and either velocity dispersion or age
\citep[e.g.][] {BernardiIV,Bernardi05} or find that the observed
scaling of Mg and H$\beta$ line profiles is consistent with
more massive ellipticals having formed earlier \citep[e.g.][]
{FFI95,FFI96}.  Moreover, the analysis of the joint correlation of
metallicity with age and velocity dispersion of
\citet{Jorgensen99,Jorgensen99b} indicates that the relation between
typical age and $\sigma$ implies very little net change in metallicity
in observed populations.  Also, it is the Mg$_{2}$ and H$\beta$
line indices which are well-correlated with velocity dispersion \citep
{Burstein88,Worthey92,Blakeslee01}; the $\left<{\rm Fe}\right>$ index
shows only weak correlation with velocity dispersion
\citep{Jorgensen97,Trager98} and so it is not clear whether this is a
result of an enhancement of $\alpha$ elements or depressed Fe, and
therefore it is difficult to translate to metallicity.  Regardless of
these uncertainties, the effect is considerably smaller than that of
changing mean spheroid ages with mass (as demonstrated in
\S~\ref{sec:ML} and \S~\ref{sec:ages} below), as e.g.\ even for the
extreme case of the evolution reported by \citet{Kuntschner00},
with $[{\rm Fe/H}]=0.56\,\log(\sigma/100\,{\rm km\,s^{-1}})$, this
results in only a change from $Z\sim0.8\,Z_{\sun}$ at
$\mbul\sim5\times10^{7}\,M_{\sun}$ to $Z \sim2.2\,Z_{\sun}$ at
$\mbul\sim2\times10^{9}\,M_{\sun}$, ultimately shifting e.g.\ the
$z=0$ B-band galaxy luminosity function by only $ \sim0.1$ magnitude.

Because these effects are weak compared to the age effects in the
stellar populations we model, we do not impose a scaling of
metallicity with mass or velocity dispersion, deferring a treatment of
the chemical enrichment histories of galaxies to future work
\citep[but see, e.g.][]{Brook04a,Brook04b,Robertson05d,Font05}, but
note that its addition does not create any conflict between our
predictions and observations.  However, these relatively small
scalings could be important for the observed colors, so we do briefly
consider the possible effects of changing metallicity with $\sigma$ in
\S~\ref{sec:colors}, where we show that the effect is small.  We do
not include the effects of dust reddening on the galaxy population, as
our simulations show a dramatic and rapid falloff in characteristic
column densities after the starburst, when the black hole expels
surrounding gas as it reaches its final mass.
This is consistent also with observations that
show that only a small fraction $\lesssim10\%$ of the luminosity in
red galaxies can come from dusty, intrinsically bluer sources
\citep{Bell04a}.

\section{The Relic Velocity Dispersion Distribution and Mass Functions}
\label{sec:sigma}

In \S~\ref{sec:sims} and \S~\ref{sec:scaling} we have determined 
$\nBH$, the rate at which quasars of a given final black hole mass are 
formed in mergers, and fit this to an analytical form.
Having also determined the $\msigma$ relation as a function of redshift and its
intrinsic dispersion from our simulations, we can then convert $\nBH$ to
$\nsigma$, the birthrate of spheroids of a given velocity dispersion
as a function of redshift.  To do so, we account for the
intrinsic dispersion of the $\msigma$ relation, by inverting
\begin{equation}
\nBH=\int^{\infty}_{0} P(\mbh | \sigma)\ \nsigma\,{\rm d}\sigma, 
\label{eqn:Nsigma}
\end{equation}
where we assume that $P(\mbh | \sigma)$ is distributed as a lognormal
about the value given by the $\msigma$ relation, with a dispersion
equal to that in our determined (and the observed) relation,
$\sim0.3$\,dex. With our modeling of spheroid and black hole co-formation in 
a single (dominant) major merger, the 
inversion of Equation~\ref{eqn:Nsigma} above is straightforward, 
as derived by \citet{YuLu04} as a method to determine the 
velocity distribution function at various redshifts for which direct observations 
of velocity dispersions are inaccessible.  
Thus, knowing
$\nsigma$, we can integrate over time (redshift) to determine the
relic number density of sources with a given velocity dispersion,
$n(\sigma) = {\rm d}n / {\rm d}\log(\sigma)$.

\begin{figure}
    \centering
    \epsscale{1.15}
    \plotone{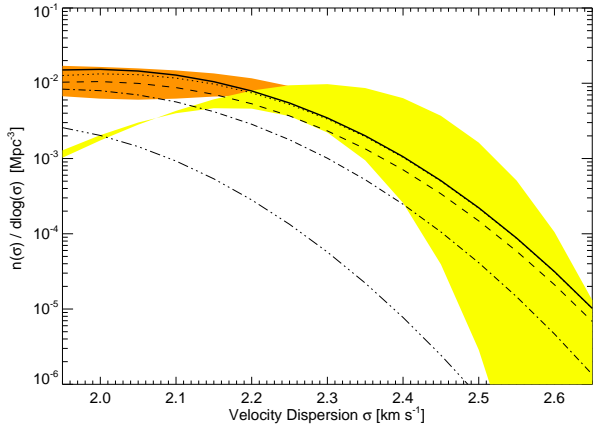}
    \caption{The relic distribution of velocity dispersions (as defined in \S~\ref{sec:scaling} and 
    as would be inferred from the $M_{\rm BH}-\sigma$ relation) at $z=0$
    (solid), 1 (dotted), 2 (dashed), 3 (dot-dashed), and 5
    (triple-dot-dashed). The $1\sigma$ range of observations of
    velocity dispersions in elliptical galaxies is shown, from
    \citet{Sheth03} (yellow shaded region), with the contribution from bulges in 
    S0 and spiral galaxies from \citet{AR02} (orange shaded region).
    \label{fig:sigma.dist}}
\end{figure}

The results of this integration to $z=0$ are shown in
Figure~\ref{fig:sigma.dist} (thick solid line).  Our theoretical
estimate agrees well with the observed distribution of velocity
dispersions found for local $z=0$ ellipticals by \citet{Sheth03}
($1\sigma$ range shown as the yellow shaded region).  The contribution
from spheroids in S0 and spiral galaxies, determined by \citet{AR02},
is added to this and shown also at the low-$\sigma$ end where it
dominates ($1\sigma$ range shown as orange shaded region).
We caution 
that our prediction at low-$\sigma$ is somewhat sensitive to the assumed faint-end 
slope in the birthrate of black holes of a given mass [$\nBH$], 
as these are not necessarily the products of major mergers. Our estimate
is on the high side at the extreme large-$\sigma$ tail of the
distribution, but this is where both the observations are uncertain
and our modeling of the quasar luminosity function and corresponding
black hole mass [$\nBH$] distribution are sensitive to the functional
form and bolometric corrections adopted.

We can also predict the velocity dispersion function at different
redshifts based on our modeling, and these results are shown in
Figure~\ref{fig:sigma.dist}.  We note that we have adopted the
pure-peak luminosity evolution (PPLE) form for the evolution of the
quasar luminosity function above $z\sim2$, where the break and
faint-end slope of the luminosity function are poorly constrained.  If
we instead consider the pure density form of this evolution, the $z=3$
and $z=5$ distributions peak at significantly higher $\sigma$.

\begin{figure}
    \centering
    \epsscale{1.15}
    \plotone{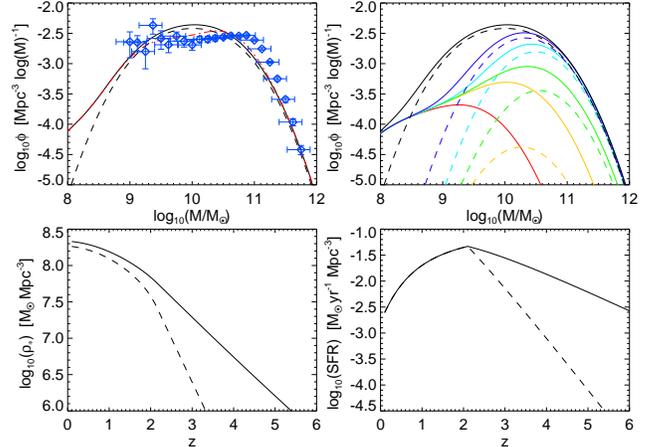}
    \caption{Predicted $z=0$ stellar mass function in remnant red,
    elliptical galaxies (upper left). This is compared to the
    morphologically selected spheroid stellar mass function of
    \citet{Bell03} (blue diamonds, where horizontal errors show the
    systematic mass uncertainty).  Red dot-dashed line shows 
    our prediction allowing for subsequent dry mergers. 
    Upper right shows the mass function
    at $z=0$ (black), $z=0.5$ (blue), $z=1$ (cyan), $z=2$ (green),
    $z=2$ (orange), and $z=5$ (red). Lower left shows the integrated
    stellar mass density as a function of redshift, lower right the
    star formation rate.  The solid lines adopt pure peak luminosity
    evolution for the quasar luminosity function above $z=2$, the
    dashed lines adopt pure density evolution.
    \label{fig:MF}}
\end{figure}

We can perform an identical procedure, using instead the relations
between black hole mass and host galaxy stellar mass to obtain the
relic stellar mass function and its evolution with redshift.
Figure~\ref{fig:MF} shows the resulting $z=0$ stellar mass function in
remnant red, elliptical galaxies (upper left). This is compared to the
morphologically selected spheroid stellar mass function of
\citet{Bell03} (blue diamonds, where horizontal errors show the
systematic mass uncertainty).  In all panels, the solid lines adopt
pure peak luminosity evolution (PPLE) for the quasar luminosity
function above $z=2$, and the dashed lines are for pure density
evolution (PDE), as defined in \S~\ref{sec:QLF}.  The agreement is
good over the entire range of observed masses, especially considering
the systematic uncertainties in the observations. As is demonstrated
for the galaxy luminosity function in Figure~\ref{fig:LF.B}, adopting
an idealized ``light-bulb'' or pure exponential light curve model for
the quasar lifetime will not produce the turnover and shallow slope of
the faint end of this mass function, and will overpredict the low-mass
end by $2-3$ orders of magnitude. The upper right of the figure shows
the mass function at various redshifts, the lower left shows the
integrated stellar mass density as a function of redshift, and the
lower right the star formation rate. The evolution of the star
formation rate qualitatively agrees well with that estimated by, e.g.\
\citet{Cole01}, but we do not account for the star-forming spiral
population which constitutes a significant or even dominant fraction
of the integrated star formation rate, and so our present results are
not necessarily in conflict with cosmological simulations indicating
that the total mean density of cosmic star formation peaks at
$z\approx 4 - 5$ (see, e.g. Springel \& Hernquist 2003b; Hernquist \&
Springel 2003; Nagamine et al. 2004a).

Subsequent gas-poor (``dry'') mergers, by definition, do not have a
reservoir of cold gas, and as a result cannot excite bright quasar
activity.  Therefore, the empirical information we derive on the rate
at which spheroids are born as a function of mass and redshift from
the quasar luminosity function does not account for dry
merging. However, we can estimate the potential impact of
spheroid-spheroid mergers on our predictions.  Recent observations
\citep{Bell05,vanDokkum05} suggest that $z=0$ spheroids have, on
average, undergone $\sim0.5-1$ major dry mergers since $z\sim0.7$ (see
also Carlberg et al.\ 1994; Le F{\`e}vre et al.\ 2000; Patton et al.\
2002; Conselice et al.\ 2003, although de Propris et al.\ 2005
estimate a significantly lower value $\sim0.2$). Observations and our
predictions for the birth redshifts of spheroids (see
\S~\ref{sec:ages}) imply that there should not be significant dry
merging much earlier, as most spheroids are either recently formed or
still forming at higher redshifts.  Therefore, we can estimate the
effects of dry merging by assuming that each spheroid has undergone
$\sim0.5$ major dry mergers in its history down to $z=0$. For
simplicity, we assume these are equal-mass dry mergers; i.e.\ for a
given interval in mass, we assume half the number of predicted
spheroids dry merge, halving their number density but doubling their
mass.

The resulting mass function is shown by the red dot-dashed line in the
upper left of Figure~\ref{fig:MF} (for the pure peak luminosity
evolution case). The net resulting change, as dry merging increases
spheroid masses but decreases the total number of spheroids, is
generally smaller than typical uncertainties in our modeling (of,
e.g.\ the functional form of $\nBH$) and the observations, and thus we
can safely ignore the impact of dry merging in our subsequent
analysis.  This is also suggested by calculation of e.g.\ the spheroid
luminosity function from semi-analytical models \citep{Cirasuolo05}.
The effect is not completely negligible, however, and we note that the
dry-merging corrected mass function agrees very well with the
observations (within $\sim1\sigma$ at all masses).  Because dry
mergers are not constrained by our empirical approach (unlike gas-rich
mergers which produce a signal in the quasar luminosity function), and
the rate and impact of dry galaxy mergers is observationally
uncertain, we do not include their effects in any of our other
predictions, but emphasize here that they introduce a relatively small
second-order effect which does not result in any conflict with the
observations.

\section{Galaxy Luminosity Functions}
\label{sec:gal.LFs}
\subsection{The B-band Luminosity Function at All Redshifts}
\label{sec:LF.B}

Unlike the relic velocity dispersion function, which is determined by
the integrated history of spheroids, the evolution of stellar
luminosities and colors makes the galaxy luminosity function in
different wavebands dependent on the time history of spheroid
formation.  Because of this, it is not implicit that successfully
reproducing the $z=0$ black hole mass distribution will guarantee an
accurate prediction for the galaxy luminosity function at $z=0$ or
higher redshifts.

As outlined in \S~\ref{sec:scaling}, we use the observed quasar
luminosity function and our simulations of quasar evolution to
determine the birthrate of black holes of mass $\mbh$, and
correspondingly spheroids of stellar mass $\mbul$, as a function of
redshift. For a given observed redshift $\zobs$, we can then integrate
over $z>\zobs$ to determine the history of the spheroids observed at
$\zobs$; i.e.\ for a given $\zobs$ and $\mbul$, the distribution of
ages/formation times is completely determined. Knowing the formation
history for these spheroids, we use the stellar population synthesis
model of \citet{BC03} to determine their observed magnitudes in any
given band at $\zobs$.

\begin{figure*}
    \centering
    \plotone{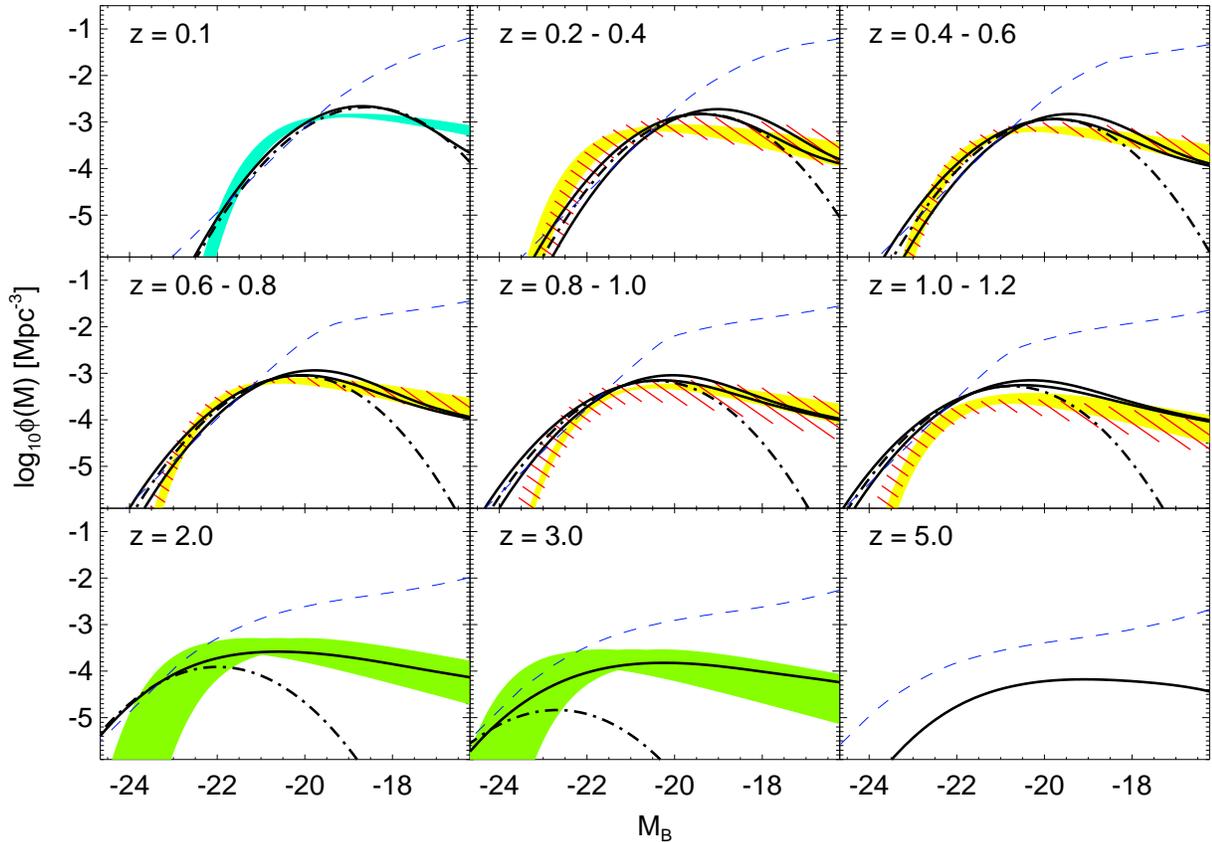}
    \caption{Predicted B-band red/elliptical galaxy luminosity
    function (solid lines) at different redshifts (shown in the upper
    left of each panel). For panels with a range of redshifts shown,
    the lines show our prediction at the minimum and maximum
    redshift. Dot-dash lines show the prediction assuming pure density
    evolution instead of pure peak luminosity evolution above $z\sim2$,
    at the mean redshift of each bin. Shaded ranges show
    the $1\sigma$ range of observed luminosity functions from
    \citet[][cyan]{Madgwick02}, \citet[][yellow and red]{Faber05},
    \citet[][green]{Giallongo05}. The blue dashed line shows the
    prediction obtained used an idealized model for the
    quasar lifetime in which quasars grow/decay exponentially or turn
    on/off as a step function.
    \label{fig:LF.B}}
\end{figure*}

We show our prediction for the rest-frame B-band red/elliptical galaxy
luminosity function at a series of observed redshifts in
Figure~\ref{fig:LF.B}. In each panel, our predicted B-band luminosity
function for the redshift indicated in the upper left is shown as the
thick black line.  When a range of $z$ is indicated in the upper left
of a panel, the predicted luminosity functions at both the minimum and
maximum redshift of the range are given. The $1\sigma$ range in the
observed luminosity function at each redshift (or redshift range) is
indicated as a shaded region. At $z\sim0-0.1$ (median $z=0.04$), the
observed luminosity function of \citet{Madgwick02}, determined from
the 2dFGRS survey, is shown in cyan. At $z=0.2-0.4,\ 0.4-0.6,\
0.6-0.8,\ 0.8-1.0,\ {\rm and}\ 1.0-1.2$, the shaded region shows the
observed luminosity functions from \citet{Faber05}, determined from
the DEEP2 (yellow) and COMBO-17 (red) surveys
\citep{Bell04b,Willmer05}.  At $z=2.0$ and $z=3.0$, the observations
from \citet{Giallongo05}, from the Hubble Deep Field and K20 surveys,
are shown in green.  At $z=5$ there is no observed B-band luminosity
function, but we show our prediction.

In each case, the observed luminosity function is determined from
either morphologically-selected elliptical galaxies or color-selected
red galaxies (especially at high redshift where morphological
information is not available), which as noted in \S~\ref{sec:intro}
are similar at least at low to moderate $z\lesssim1-2$ redshifts
\citep[e.g.][]{Strateva01,BernardiIV,Bell04a,Ball05}.  Our predictions
agree well with the observations, over a wide range of magnitudes and
redshifts.  We slightly overpredict the bright end of the luminosity
function at high redshift, but this can be explained by selection
effects, as we show below in \S~\ref{sec:colors}, because many of
these very bright, high-redshift galaxies are quite blue (as they have
formed only recently at these high redshifts) and thus would not
appear in an observed red galaxy luminosity function (although this is
also somewhat related to our slight overprediction of the
high-$\sigma$ end of the velocity dispersion function in
Figure~\ref{fig:sigma.dist}).

In Figure~\ref{fig:LF.B}, we also show the predicted B-band
red/elliptical galaxy luminosity function at each redshift using a
commonly employed, idealized model for the quasar lifetime (blue
dashed lines).  Here, we assume that a quasar radiates at its peak
luminosity $L=\Lp$ for a fixed time equal to $10^{7}$\,yr (as is often
adopted, and similar to the Salpeter time for $e$-folding of an
Eddington-limited black hole, $t_{S}=4.2\times10^{7}$\,yr), but we
note that the entire class of ``light-bulb'' or exponential
growth/decay models for the quasar light curve produces a nearly
identical prediction to that shown.  This model overpredicts the
number of red/elliptical galaxies which should be observed at low
luminosities by two orders of magnitude, does not reproduce the shape
and curvature of the luminosity function, and underpredicts the bright
end if the lifetime is chosen to be longer (e.g.\ the actual Salpeter
time). The quasar lifetime in these models is a free parameter, but it
determines only the normalization of this curve, and thus no value can
produce a reasonable prediction for the galaxy luminosity function.
The reason for the failure of these models at low luminosity is, as
mentioned above, the fact that they associate objects observed at low
luminosities with low-$\Lp$ objects, and therefore low-$\mbh$ objects
in small-mass spheroids.

Figure~\ref{fig:LF.B} also shows our prediction (dot-dashed lines),
for the mean redshift of each bin, assuming pure density evolution
(PDE) instead of pure peak luminosity evolution (PPLE) for the
birthrate of quasars with a given peak luminosity above $z\sim2$.
Although the observed quasar luminosity function does not provide a
good constraint on which evolution is followed, the difference in our
subsequent calculations is usually minimal, and observations of the
faint end of the galaxy luminosity function at moderate and high
redshifts (where the two predictions begin to diverge) do not yet
exist. However, if such observations of the galaxy population can be
made, or the ages of the lowest-mass/luminosity objects at $z\sim0$
are measured, they can provide a powerful constraint on the $\nLp$ [$\nBH$,
$\nbul$] distributions (i.e.\ the rates at which spheroids and quasars
of given properties form with redshift).

\subsection{The Evolution of the Luminosity Function with Redshift}
\label{sec:LF.B.evol}

The observed galaxy luminosity function is usually fit to a Schechter
function \citep{Schechter76} with normalization $\phistar$,
characteristic magnitude (luminosity) $\Mstar{}$ ($L^{\ast}$), and
faint-end slope, $\alpha$. This yields a total number density of
galaxies $\Phi=\phistar\,\Gamma(\alpha+1)$, and a total luminosity
density $j=\phistar\,L^{\ast}\,\Gamma(\alpha+2)$.  We can determine
$\Phi$ and $j$ by integrating our predicted luminosity function at
each redshift. However, observationally, it is easier to determine
$\phistar$ than $\Phi$, as $\alpha$ is difficult to measure and a
constant $\alpha$ is often assumed.  To compare directly with most
observations
\citep[e.g.,][]{Cohen02,Bell03,Madgwick03,Giallongo05,Faber05}, we
therefore assume a constant $\alpha_{0}=0.5$ and calculate
$\phistar\equiv\Phi/\Gamma(\alpha_{0}+1)$ and likewise calculate
$\Mstar{}$ [$L^{\ast}\equiv j/(\phistar\,\Gamma(\alpha_{0}+2))$].

\begin{figure*}
    \centering
    \epsscale{1.15}
    \plotone{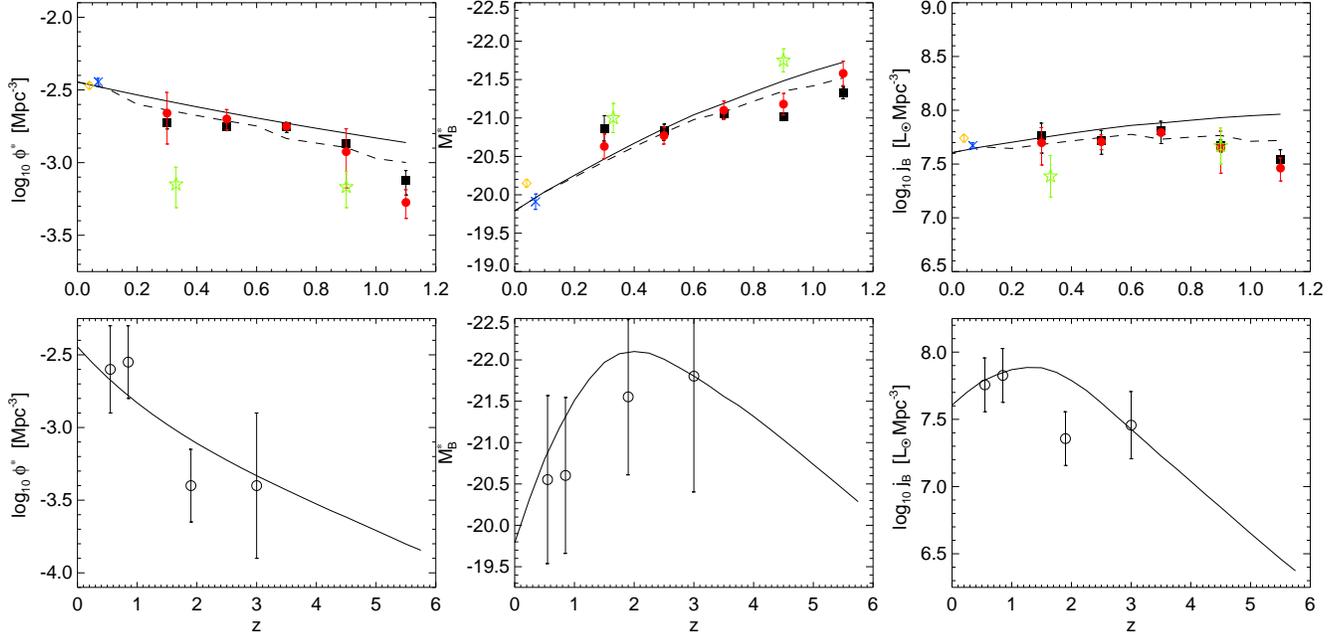}
    \caption{B-band luminosity function normalization $\phistar$
    (left), characteristic magnitude $\Mstar{B}$ (center), and
    luminosity density $j_{B}$ (right) predicted by our model (black
    lines) as a function of redshift for $z=0-6$. Upper panels show
    the $z\lesssim1$ range in greater detail. Dashed lines show the 
    prediction ignoring recent ($<1$\,Gyr) mergers. Observations are from
    \citet{Faber05} (COMBO-17, red circles, and DEEP2, black squares),
    \citet{Madgwick03} (2dF, orange diamonds), \citet{Bell03} (SDSS,
    blue $\times$'s), and \citet{Im02} (DEEP1, green stars) for the
    low-redshift (upper) panels. Results from \citet{Giallongo05}
    (Hubble Deep Field and K20, open black circles) at high redshift
    (lower panels) are also shown. 
    \label{fig:LF.B.evol}}
\end{figure*}

Figure~\ref{fig:LF.B.evol} shows $\phistar$, $\Mstar{B}$, and $j_{B}$
as a function of redshift.  Our prediction is shown as a solid black
line both in a low redshift interval $z<1.2$ (upper panels) and over
the entire $z<6$ interval (lower panels).  At low redshifts (upper
panels), observations from \citet{Faber05} (COMBO-17, red circles, and
DEEP2, black squares), \citet{Madgwick03} (2dF, orange diamonds),
\citet{Bell03} (SDSS, blue $\times$'s), and \citet{Im02} (DEEP1, green
stars) are shown, with $1\sigma$ errors, and at high redshifts (lower
panels), observations from \citet{Giallongo05} (Hubble Deep Field and
K20) are shown (circles).

Although we slightly overpredict $\phistar$ (and thus $j_{B}$ as a
consequence) at $z\sim1.2$, this is related directly to our small
overprediction of the bright blue end of the luminosity function
discussed in \S~\ref{sec:LF.B} and, as discussed in
\S~\ref{sec:colors} can be explained by selection effects as these
objects have recently formed and are bluer than their traditionally
color-selected counterparts. We estimate the results of this selection
effect in the upper panels, where the dashed lines show our prediction
ignoring all objects which have formed (i.e.\ gone through their peak
merger/quasar activity) less then 1 Gyr in the past, and thus have not
had sufficient time to redden to the point where they would be
recognized as red galaxies in color-selected surveys (this corresponds
roughly to the color selection of e.g.\ Bell et al.\ 2004b, given our
modeled metallicities and star formation histories).  The agreement at
$z\sim1-1.2$ is significantly improved, suggesting that the strong
increase in red galaxies from $z\sim1$ to present is driven in part by
continued formation and mergers associated with ongoing (though
declining) quasar activity, and in part by the reddening of spheroids
formed in mergers at the peak of quasar activity $z\sim1-2$, reddening
to the point where they will be recognized as red ellipticals by
$z\sim0$.

Because, in our picture, spheroids and quasars form together through
mergers, the quantities $\phistar$, $\Mstar{B}$, and $j_{B}$ are
directly related to the quasar luminosity function.  Associating each
merger with a single quasar and spheroid, the total number of red
galaxies is given by the integrated number of quasars produced
up to the observed redshift; i.e.\
$\Phi=\int_{\infty}^{\zobs}\dot{n}(QSO)\,dt$, where $\dot{n}(QSO)$ is
the number density of quasars born per unit time per unit comoving
volume. In our determination of the luminosity function, this is
$\dot{n}_{\ast}=\,$constant, the normalization of the lognormal $\nBH$
distribution.  Thus $\phistar=\dot{n}_{\ast}
t_{H}(z)/\Gamma{(\alpha+1)}$, where $t_{H}$ is the age of the Universe
at a particular redshift. Note that if we adopted pure density
evolution for the quasar luminosity function above $z\sim2$,
$\dot{n}_{\ast}$ would fall off exponentially above these redshifts,
and $\phistar$ would drop correspondingly. Currently, the observations
are insufficient to decide which possibility is correct, but this
makes it clear that estimating the total number of red galaxies
at high redshift in future observations can constrain the form of the
quasar luminosity function evolution.

Likewise, $\Mstar{}$ is directly related to the break in the observed
{\it quasar} luminosity function, which in turn corresponds directly
to the peak in the $\nLp$ [and corresponding $\nBH$] distribution
\citep{H05c}, and thus gives the peak in the rate at which spheroids
of a given stellar mass are forming as a function of that stellar
mass. Because luminosities evolve with the age of the stellar
population, this is not trivially related to the $\Mstar{}$ of the
galaxy population as $\Phi$ is to the number density of quasars being
formed, but the two are still critically related and, in general,
increasing $\Mstar{}$ corresponds to moving the break in the observed
quasar luminosity function to higher luminosities, and vice versa.

\subsection{The Luminosity Function in Different Wavebands}
\label{sec:LF.bands}

\begin{figure}
    \centering
    \epsscale{1.2}
    \plotone{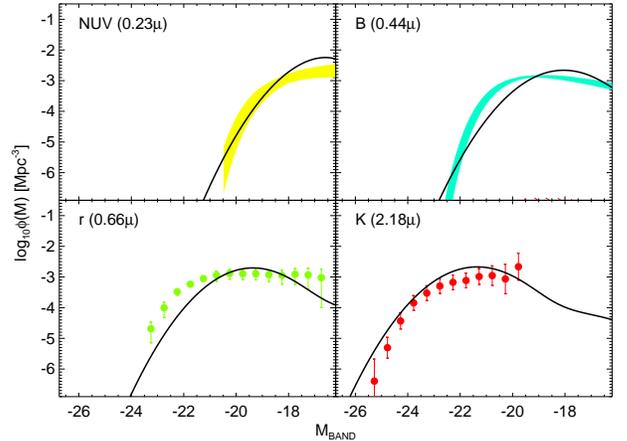}
    \caption{Predicted $z\sim0$ red/elliptical galaxy luminosity
    function in four wavelengths (black lines), in the manner of
    Figure~\ref{fig:LF.B}. Observations are from
    \citet{Budavari05,Treyer05} in the near UV (NUV; yellow; upper
    left), \citet{Madgwick02} in B-band (cyan; upper right),
    \citet{Nakamura03} in Sloan $r$-band (green; lower left), and
    \citet{Kochanek01} in K-band (red; lower right).
    \label{fig:LF.bands.z0}}
\end{figure}

Figure~\ref{fig:LF.bands.z0} shows our predicted red/elliptical galaxy
luminosity function (solid lines) in several different wavebands at
$z\sim0$; the near ultraviolet (NUV; at $2400$\,\AA\ or $0.24\,\mu$),
B-band ($0.44\,\mu$), r-band ($0.66\,\mu$), and K-band
($2.18\,\mu$). Each is compared to the observations (shaded regions or
points showing $1\sigma$ errors), shown over the range of magnitudes
where data exist. The observations shown are from \citet{Budavari05}
and \citet{Treyer05} in the NUV from GALEX (yellow; upper left),
\citet{Madgwick02} in B-band from 2dFGRS (cyan; upper right),
\citet{Baldry04} (see also Nakamura et al.\ 2003) in Sloan $r$-band
from SDSS (green; lower left), and \citet{Kochanek01} in K-band from
2MASS (red; lower right). The NUV prediction has been rescaled to AB
magnitudes for ease of comparison with the observations.  The
agreement in these bands is good, implying that not only do we
reproduce the luminosity function in a wide variety of wavebands, but
also the color distribution as a function of magnitude.  

\begin{figure}
    \centering
    \epsscale{1.2}
    \plotone{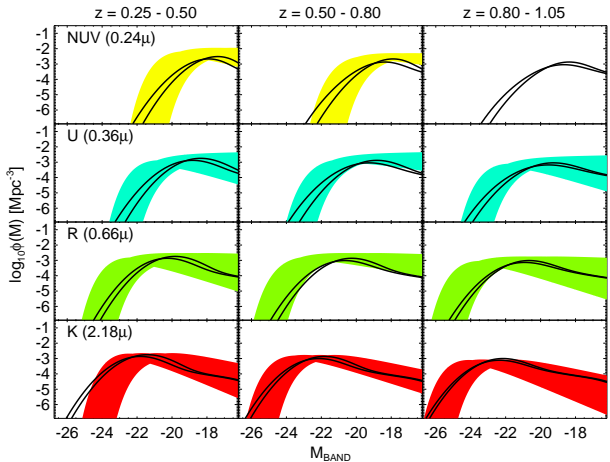}
    \caption{Predicted luminosity function (black lines) at the
    minimum and maximum redshift of each redshift range shown,
    $z=0.25-0.5$ (left panels), $z=0.5-0.8$ (center panels), and
    $z=0.8-1.05$ (right panels), in the manner of
    Figure~\ref{fig:LF.B}. Each is compared to observations from
    \citet{Cohen02} (except the NUV at $z=0.8-1.05$, where the
    observations are insufficient to determine a luminosity function),
    in the NUV (yellow), U (cyan), R (green), and K (red) bands.
    \label{fig:LF.bands.difz}}
\end{figure}

Figure~\ref{fig:LF.bands.difz} extends this to higher redshift,
showing the predicted luminosity function in the NUV (yellow, top
panels), U-band ($0.36\,\mu$; blue, second from top), R-band (green,
second from bottom), and K-band (red, bottom panels) in three redshift
intervals, from \citet{Cohen02}. Again, the shaded regions show the
$1\sigma$ range in the observed luminosity function and the solid
lines show our prediction at the minimum and maximum redshift of each
interval. Our predictions also agree well with the VIMOS luminosity
functions in U, B, V, R, and I from \citet{Zucca05} for the redshift
range $z=0.4-0.9$ (these results compare favorably with the plotted
luminosity functions in the center panels).

\begin{figure*}
    \centering
    \plotone{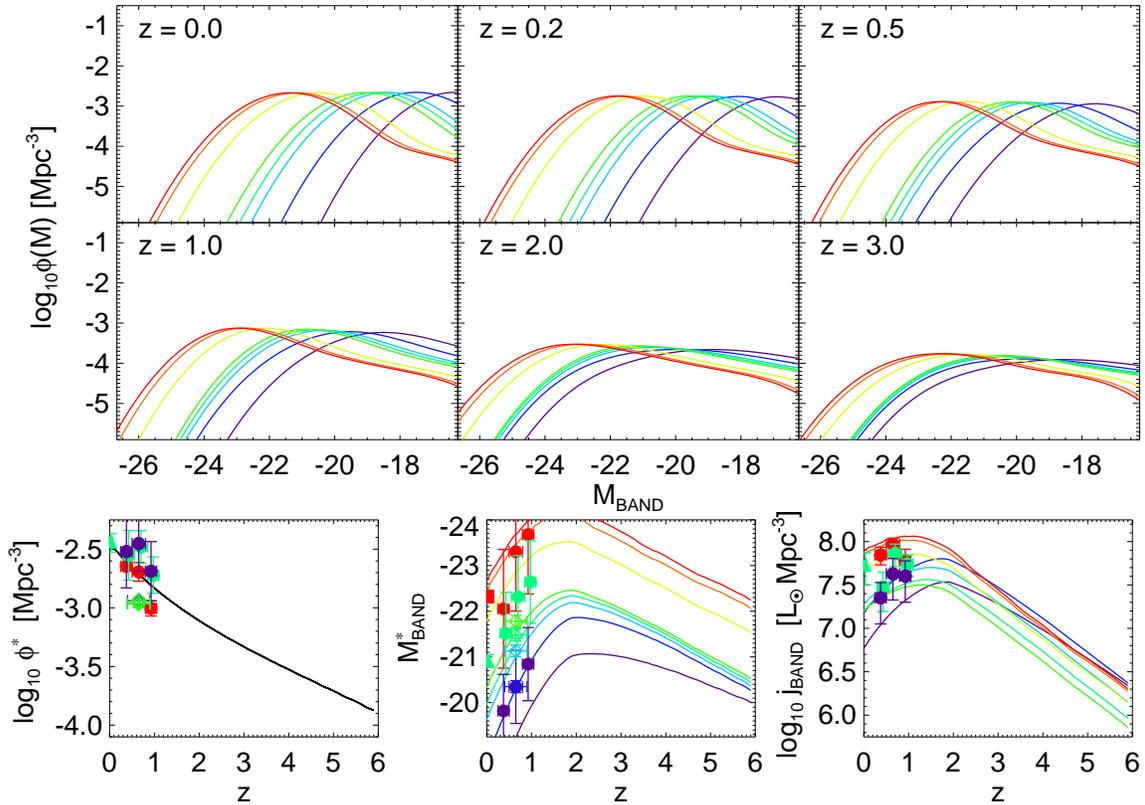}
    \caption{Predicted luminosity function at six representative
    redshifts (upper left of each panel) in several bands: U (purple),
    B (blue), V (cyan), R (light green), I (green), J (yellow), H
    (orange), and K (red) (with generally decreasing characteristic
    magnitude). Lower panels show $\phistar$, $\Mstar{BAND}$, and
    $j_{\rm BAND}$ for each band, in the manner of
    Figure~\ref{fig:LF.B.evol}. Observed points for $\phistar$,
    $\Mstar{BAND}$, and $j_{\rm BAND}$ in U, R, and K bands are shown
    (filled circles of the appropriate color) from \citet{Cohen02}, at
    $z=0$ from \citet{Kochanek01} (K, red square) and
    \citet{Nakamura03} (r, green triangle), and at $z=0.4-0.9$ from
    \citet{Zucca05} (U, B, V, R, I, diamonds of appropriate colors).
    \label{fig:LF.allbands}}
\end{figure*}

In Figure~\ref{fig:LF.allbands}, we plot the predicted luminosity
function at redshifts $z=0.0,\ 0.2,\ 0.5,\ 1.0,\ 2.0,\ {\rm and}\
3.0$, and $\phistar$, $\Mstar{BAND}$, and $j_{\rm BAND}$ (the
normalization, characteristic magnitude, and total luminosity density
in each band, respectively) of each luminosity function (determined as
in \S~\ref{sec:LF.B.evol}) for redshifts $z=0-6$. The results are
shown for the bands U, B, V, R, I, J, H, and K, from purple to red,
respectively. For $\phistar$, $\Mstar{BAND}$, and $j_{\rm BAND}$, the
U, R, and K-band observations of \citet{Cohen02} (from the luminosity
functions of Figure~\ref{fig:LF.bands.difz}) are shown as filled
circles (with colors matching those of the corresponding prediction
for each band). The $z=0.4-0.9$ observations in U, B, V, R, I (with
the corresponding colors) from \citet{Zucca05} are shown also
(diamonds), as are the $z\sim0$ observations of \citet{Nakamura03} (r,
green triangle) and \citet{Kochanek01} (K, red square).  This provides
a large set of predictions, of the shape and integrated properties
($\phistar,\ \Mstar, j$) of the red galaxy distribution, for future
comparison with red or elliptical galaxy luminosity functions.

\section{The Color Distribution of Red Galaxies as a Function of Magnitude and Redshift}
\label{sec:colors}
\nobreak

\begin{figure*}
    \centering
    \plotone{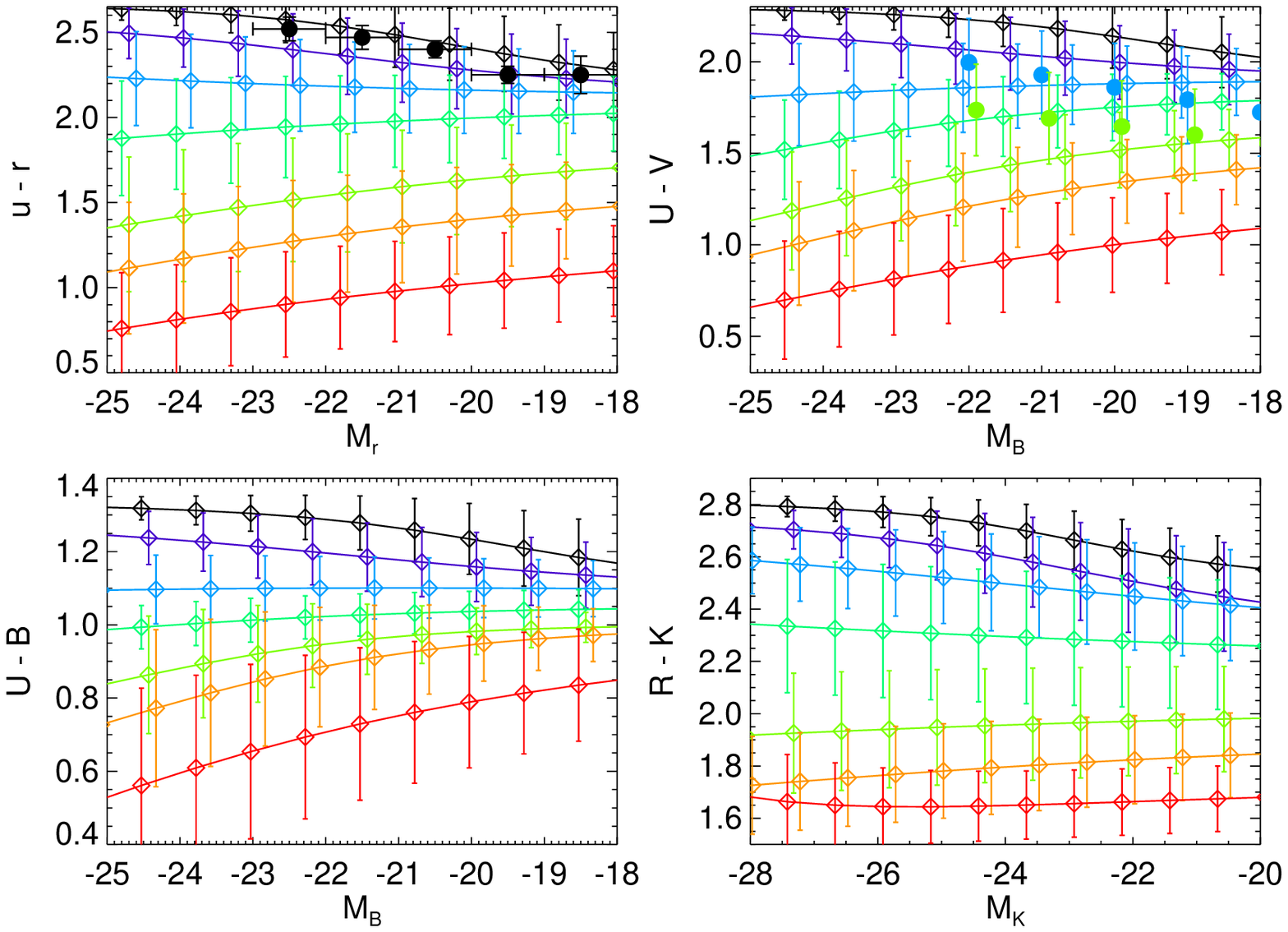}
    \caption{Predicted mean color (diamonds and lines) as a function
    of magnitude for several color-magnitude pairs (as labeled), with
    rms deviation in the color distribution (vertical
    errors). Our predictions are shown for $z=0$ (black), $z=0.2$
    (purple), $z=0.5$ (blue), $z=1$ (cyan), $z=2$ (green), $z=3$
    (orange), $z=5$ (red), with bluer colors at higher redshift. Our
    results are compared to observations of $(u-r)$ vs.\ $M_{r}$ at $z=0$
    \citep[black circles, upper left;][]{Balogh04}, and
    $(U-V)$ vs.\ $M_{B}$ from $z=0.4-1.0$ and $z=1.3-3.5$ \citep[blue and
    green circles, respectively, upper right;][]{Giallongo05}.
    Pure density evolution is assumed for the quasar luminosity 
    function above $z\sim2$.
    \label{fig:color.mag}}
\end{figure*}

Figure~\ref{fig:color.mag} shows our predicted color-magnitude
relations for several different wavebands at a series of
redshifts.
We plot the mean colors (lines and open diamonds) at each
magnitude and redshift, with the rms dispersion in the color
distribution shown as vertical error bars.  We show four separate
color-magnitude diagrams, for comparison with a range of
observations. These are $(u-r)$ vs.\ $M_{r}$ (upper left), as observed
in e.g., Baldry et al. (2004) and Balogh et al. (2004), $(U-V)$ vs.\
$M_{B}$ \citep[upper right;][]{Cross04,Giallongo05,McIntosh05a},
$(U-B)$ vs.\ $M_{B}$ \citep[lower left;][]{Willmer05,Faber05}, and
$(R-K)$ vs.\ $M_{K}$ \citep[lower
right;][]{Roche02,Pozzetti03,Fontana04}.
For $(u-r)$ vs.\ $M_{r}$, we show
the $z=0$ color-magnitude relation determined by \citet{Balogh04} as
solid black circles, with corresponding errors.
We also show the observed $(U-V)$ vs.\
$M_{B}$ color-magnitude relations (filled circles) at $z=0.4-1.0$
(blue) and $z=1.3-3.5$ (green) from \citet{Giallongo05}, and find
reasonable agreement despite the much larger uncertainties at these
larger redshifts.  The $z\sim0$ determination of $(U-V)$ vs.\ $M_{B}$
from \citet{McIntosh05a} also agrees with our prediction.
 
We note that although our predicted $(R-K)$ colors are not as red as
those of extremely red objects observed at high redshift
\citep[e.g.,][]{Roche02,Franx03}, we are not attempting to reproduce
this population, which is heavily influenced by the presence of
ongoing starbursts and dust reddening, and possible AGN activity as is
typical of e.g.\ low-redshift ultraluminous infrared galaxies
\citep[e.g.,][]{Roche02,Miyazaki03,SM96}. Our predictions are,
however, consistent with the $(R-K)$ colors of ellipticals observed
by, e.g., Pozzetti et al.\ (2003). The presence of even mild dust
reddening, which we do not expect to have a large impact on most of
the colors and magnitudes we show, based on the rapid falloff in
column densities post-merger \citep{H05a}, will, however,
strongly redden the $(R-K)$ colors. It is therefore not surprising
that our predicted, intrinsic, non-dust reddened $(R-K)$ colors are
too blue, and this demonstrates that reproducing these colors will
require more sophisticated models which incorporate dust reddening in
the ISM and possibly the continued production of dust in stellar
winds.

Our modeling reproduces the observed color-magnitude relations of
red/elliptical galaxies over the range of magnitudes observed and for
different observed colors. Furthermore, the typical dispersion about
the mean color at low redshift, $\sim0.2$, agrees well with that
observed for this population of galaxies \citep{Baldry04,Balogh04}.
We predict the evolution in this dispersion with redshift, in good
agreement with \citet{vanDokkum00}, who find based also on the
observations of \citet{Bower92}, \citet{Ellis97}, and
\citet{vanDokkum98} that the scatter in the color-magnitude [$(U-B)$
vs.\ $M_{B}$, specifically] relation of all progenitors of present
early-type galaxies increases by a factor $\sim2$ between $z=0$ and
$z=1$.  Moreover, we reproduce the observed trend of increasingly
blue colors at higher redshift
\citep[e.g.,][]{Bell04b,Cross04,Giallongo05} as these galaxies have
formed more recently and thus not reddened as much.  This is clear
from the comparison with the observations of \citet{Balogh04} and
\citet{Giallongo05} shown, but further, the observed ``blueing'' of
the red galaxy population is observed to be $\sim0.3$ magnitudes over
the redshift range $z\sim0-1$ \citep{Bell04b}.

\begin{figure}
    \centering
    \epsscale{1.15}
    \plotone{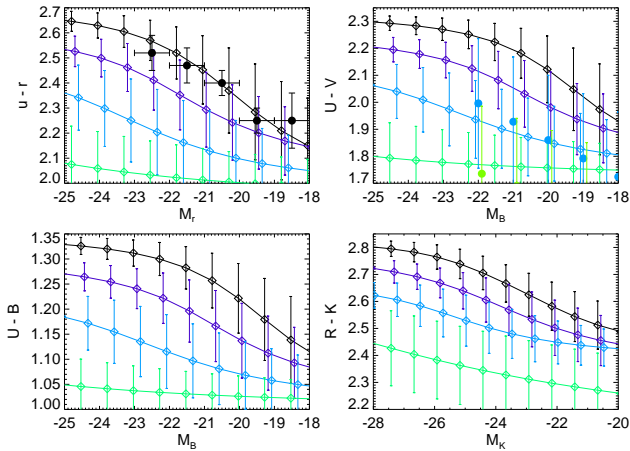}
    \caption{Same as Figure~\ref{fig:color.mag}, but excluding all spheroids 
    which have formed less than 1 Gyr before the observed redshift. 
    \label{fig:color.mag.old}}
\end{figure}

At high redshifts shown in Figure~\ref{fig:color.mag}, the slope of
the color-magnitude relation changes, and brighter objects become
bluer than fainter ones. The magnitude of this change in slope depends
on whether we adopt a pure peak luminosity evolution (PPLE) or pure
density evolution (PDE) form for the quasar evolution at high
redshifts, as shown below in Figure~\ref{fig:color.mag.tracks}.
Beyond this, however, this change in slope and normalization owes to
the fact that the most massive remnant galaxies form at redshifts
$z\sim2$, corresponding to the observed peak in bright quasar activity
generated in mergers. Thus, at high redshift, these objects have
formed more recently, and are bluer.

There is some evidence for this, as, e.g., \citet{Giallongo05} find a
$\sim30\%$ change in the slope of the $(U-V)$ vs.\ $M_{B}$ relation
from $z=0.4-1$ to $z=1.3-3.5$, consistent with our predictions.
Still, although the observations do not strongly distinguish between
the PPLE and PDE cases at this point, the weaker slope evolution seen
in the PDE case is somewhat more consistent with the observations of
\citet{vanDokkum00}, \citet{Bower92}, \citet{Ellis97}, and
\citet{vanDokkum98}, who find results consistent with no evolution in
the $(U-B)$ vs.\ $M_{B}$ slope at redshifts $z=0-1$, and at most a
similar $\sim30-40\%$ change over this redshift range.  However, we
caution that these samples are selected either by color (in which case
they are obviously biased against a strong blueing of the high-mass
population) or by morphology. If a considerable fraction of the most
massive galaxies are still forming (i.e. have recently merged or begun
merging), they will not have relaxed and will not be identified by
either criterion.  Therefore, we consider the color-magnitude relation
derived if we ignore all objects at any redshift which have formed
less than 1 Gyr in the past (about the time it takes for significant
morphological and color disturbances from the merger to relax).

Figure~\ref{fig:color.mag.old} shows our predictions with this caveat
(in the manner of Figure~\ref{fig:color.mag}, also assuming pure
density evolution above $z\sim2$), for $z=0-1$, as at higher redshifts
this cut excludes all but the objects formed at the highest, most
uncertain redshifts. As is clear in the figure, this further reduces
the evolution in the slope, with the change in slope over this
redshift range in each color magnitude relation essentially consistent
with zero.

We do not explicitly model populations of ``old'' pre-merger stars
(although these are included in our simulations), which should form in
the progenitor disks before the merger.  Although at times long after
the merger this should not be a significant contributor to the galaxy
colors, as much of the stellar population is formed in a strong
starburst, the effect could be significant for massive galaxies which
have recently formed, reddening these objects and reducing (or even
reversing) the slope evolution shown.  Regardless, this slope change
is difficult to observe, even in the absence of the strong limits to
measured magnitudes and colors imposed from observations at higher
redshift, as some of these objects become blue or morphologically
disturbed enough that they will not be classified as red/elliptical
galaxies.  This explains our slight overprediction of the very bright
end of the galaxy luminosity function at redshifts $z\gtrsim1-2$ in
\S~\ref{sec:LF.B} and \S~\ref{sec:LF.B.evol}, as these galaxies
correspond to the rapidly blueing galaxies in these color-magnitude
relations and will not appear in the observed red galaxy luminosity
functions.  

As noted above, we also do not include the effects of dust
reddening, which can become important for recently-formed galaxies in
which star formation has not yet terminated (i.e.\ massive galaxies at
high redshift), as our modeling in \citet{H05a} and observations of
the high-redshift massive red galaxy populations \citep{Labbe05}
indicate, and will most likely also reduce or even reverse the plotted
evolution in slope. However, we do not expect this to have a strong
effect on the typical mean colors at a given redshift, except perhaps
for the very highest redshifts where most galaxies may still be
actively merging.

Despite these caveats, we can make two further predictions from our
modeling.  First, the observed bimodality in the distribution of
galaxy colors should break down at large redshift, especially at high
luminosities, as the bright-end merger remnants become
bluer. Specifically, we predict, in the absence of strong evolution in
the blue color population, that the two color distributions should
coalesce around $z\sim1.5-2$, as is observed by, e.g., Willmer et
al. (2005) and Giallongo et al. (2005).  Second, the fraction of red
galaxies (classified on the basis of the $z\sim0$ bimodal color
distribution), which dominate the bright end of the luminosity
function at low redshift, should decrease at higher redshift (i.e.\
the bright end of the luminosity function should have an increasing
contribution from ``blue'' galaxies, in reality the same as the red
elliptical remnants observed at $z\sim0$ but formed more recently and
thus bluer), as observed by \citet{Cross04}, \citet{Daddi04}, and
\citet{Somerville04}.  These authors find a fraction as large as
$\sim1/3-1$ of these galaxies show irregular morphologies providing
evidence for merger-driven interactions by $z\sim1.5-2.0$, as we
expect based on their formation redshifts (see also
Figures~\ref{fig:ages} and \ref{fig:frac.young} below).  This also
explains the observations of \citet{Arnouts05} in the far UV
($1500\,$\AA) from GALEX and \citet{Brinchmann98} in HST morphological
surveys, who find that the density of unobscured starburst or peculiar
merging galaxies increases dramatically from $z=0$ to $z\sim1$, where
they begin to dominate the bright end of the cumulative (spiral and
elliptical) luminosity function, as anticipated from the
color-magnitude evolution of Figure~\ref{fig:color.mag} and the excess
of bright blue (recently forming) galaxies beginning to appear at this
redshift in Figure~\ref{fig:LF.B}.  This is also expected from our
modeling of the co-production of quasars and spheroids, as numerous
observations have found that the host galaxies of quasars at high
redshift (which should relax to become normal present ellipticals) are
excessively blue, both from AGN contributions and recent starburst
activity \citep[see, e.g.][and references therein]
{Bahcall97,CS01,Dunlop03,Sanchez04,Jahnke04}.  Furthermore,
\citet{Labbe05} find that dusty blue galaxies which are still forming
stars constitute a large fraction ($\sim70\%$) of the high-mass red
galaxy population at $z\gtrsim2-3$, while older ``dead'' red spheroids
constitute a smaller fraction $\sim30\%$, with ages implying formation
redshifts $z\lesssim5$ \citep[accounting for a rapid quenching of star
formation instead of ongoing star formation, see
e.g.][]{FS04,vanDokkum04}.

\begin{figure*}
    \centering
    \plotone{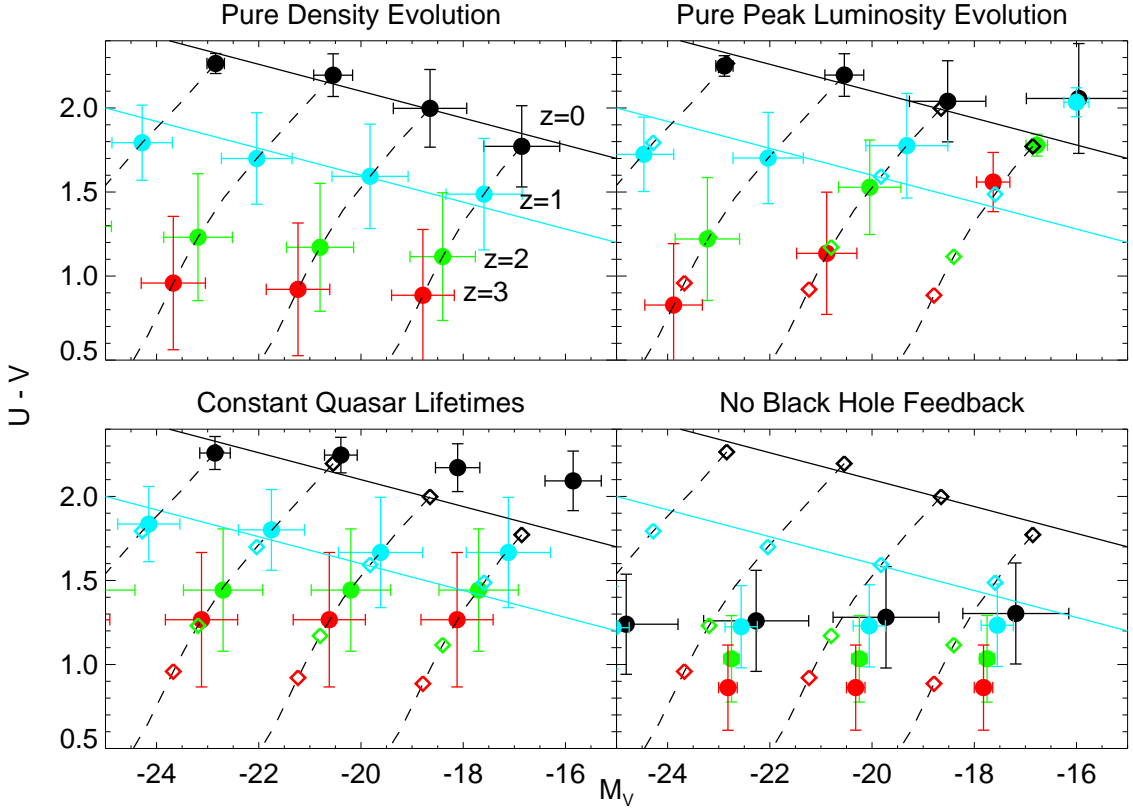}
    \caption{Predicted mean $(U-V)$ color and $M_{V}$ magnitude (circles) with 
    rms dispersion in color and magnitude (vertical and horizontal errorbars, respectively) 
    as a function of redshift at $z=0$ (black), $z=1$ (blue), $z=2$ (green), and $z=0$ (red), 
    for galaxies with total stellar mass $\mbul=10^{9},\ 10^{10},\ 10^{11},\ {\rm and}\ 
    10^{12}\,M_{\sun}$, from right to left respectively. 
    Our standard modeling, assuming pure density evolution (PDE) in the quasar population 
    above $z\sim2$, is shown in the upper left, with dashed lines showing the full 
    color-magnitude tracks from $z=0$ to $z>6$. The dashed lines and PDE points from the 
    upper left are reproduced in the other panels (diamonds), which show the mean color 
    and magnitude with redshift assuming pure peak luminosity evolution above $z\sim2$ 
    (upper right), adopting a constant quasar lifetime or exponential quasar light curve 
    (lower left), or ignoring black hole feedback in mergers (lower right). Solid line show the 
    observed color-magnitude relations at $z=0$ \citep{Bower92,SS92,Terlevich01} 
    and $z=1$ \citep{Bell04b,Giallongo05} in black and blue, respectively. 
    \label{fig:color.mag.tracks}}
\end{figure*}

Figure~\ref{fig:color.mag.tracks} shows the color-magnitude [$(U-V)$
vs.\ $M_{V}$ shown] tracks with redshift, for the population of
spheroids of fixed total stellar mass $\mbul=10^{9},\ 10^{10},\
10^{11},\ {\rm and}\ 10^{12}\,M_{\sun}$, from right to left,
respectively (i.e.\ decreasing magnitude with increasing stellar
mass).  In the upper left, we show (dashed lines) the tracks predicted
by our modeling, assuming pure density evolution for the quasar
luminosity function above $z\sim2$, from the bluest colors below the
range plotted at $z\gtrsim6$ to the reddest colors at $z=0$. The
tracks show the mean color and magnitude of the population of objects
at the given mass, as observed at a given redshift.  For comparison,
we also plot the observed $z=0$ [black;
$(U-V)\approx2.1-0.08\,(M_{V}+20)$] \citep{Bower92,SS92,Terlevich01}
and $z=1$ (blue; same slope but normalization lower by $\sim0.4$\,mag)
\citep{Bell04b,Giallongo05} color-magnitude relations as solid lines.

The agreement with the observed color-magnitude relations is good.  At
high redshift, galaxies of all masses are still forming, and so the
mean colors are blue, and there is no significant slope in the
color-magnitude diagram. However, the peak of bright quasar activity
at $z\sim2-3$ corresponds to the peak in the formation of massive
spheroids via gas-rich mergers (subsequent dry merging does not affect
our results).  Feedback from black hole growth quenches further star
formation following a merger, and the massive remnants quickly
redden. However, the typical spheroids being formed shift to lower
masses, as quasars evolve to smaller characteristic luminosities with
decreasing redshift, keeping the population blue at lower masses, and
yielding the slope of the color-magnitude diagram. This illustrates
the anti-hierarchical growth of both the black hole and spheroid
populations, and their self-consistency given our model of quasar
lifetimes to connect the two populations.

In the upper right of Figure~\ref{fig:color.mag.tracks}, we show the
theoretical result assuming pure peak luminosity evolution (PPLE) in
the quasar population above $z\sim2$, and reproduce the pure density
evolution (PDE) tracks (dashed lines) and points at redshifts $z=0,\
1,\ 2,\ 3$ (diamonds) for comparison. At low redshifts, the agreement
with observations is similar.  While there is a discrepancy at the
lowest masses $\mbul=10^{9}\,M_{\sun}$, this is both where the
observations are uncertain and where our prediction is sensitive to
the form of the faint-end $\nLp$ [$\nBH$] distribution adopted, and,
within observational uncertainty, can be slightly adjusted to yield
agreement with the $z=0$ color-magnitude relation at these low masses.
The evolution in the slope of the color magnitude relation is stronger
in the PPLE case than the PDE case because, above $z\sim2$, the PDE
model predicts a distribution in formation rates that decreases
uniformly with redshift, implying that objects of any given mass at
these redshifts have the same {\em fractional} population from earlier
redshifts.  However, the PPLE case assumes that the distribution of
formation rates shifts to lower luminosities above $z\sim2$ rather
than uniformly decreasing, implying that before $z\sim2$, most of the
lowest mass objects were formed earliest while larger objects only
just formed, with this trend reversing subsequently.  Because most
spheroid and quasar production occurs after $z\sim2-3$, this is
sufficient to reproduce the observed $z=0$ relations, but results in
the stronger slope evolution, even a reversal in sign in the
color-magnitude relation slope at high redshifts.  Therefore, our
probes of the mean ages and in particular the age distribution of even
low-redshift low-mass spheroids, as well as the color-magnitude
relation at moderate and large redshifts, can constrain the evolution
in the {\em high-redshift quasar population}.

In the lower left of Figure~\ref{fig:color.mag.tracks}, we show the
prediction (in the same manner as the upper right panel, again
reproducing the upper left panel results of our standard modeling for
comparison), assuming a constant quasar lifetime, exponential,
or ``on/off'' model of the quasar light curve.  The exact value of the
quasar lifetime we chose is unimportant, as it sets only the
normalization of the number of spheroids produced, not their
magnitudes or color distribution.  It is clear that such a model does
not accurately reproduce the $z=0$ color-magnitude relation, even at
moderate spheroid masses $\mbul\sim10^{10}-10^{11}\,M_{\sun}$.  This
is because such modeling does not incorporate strong enough `cosmic
down-sizing'; i.e.\ a sufficiently strong age gradient with spheroid
mass, even allowing for a quasar luminosity function with strong
``luminosity-dependent density evolution'' as e.g.\ the Ueda et al.\
(2003) luminosity function adopted here.

The lower right panel shows our predicted color-magnitude diagram
neglecting black hole feedback in galaxy mergers. As demonstrated by
\citet{SDH05a}, mergers without black hole feedback result in much
weaker heating of the gas in the galaxy, so that star formation
continues, declining in a roughly exponential manner over a Hubble
time, as found in simulations without black holes by e.g.\
\citet{MH94,MH96}.  Therefore, we can approximate the prediction in a
model neglecting black hole feedback by allowing for an exponentially
declining star formation rate after a peak corresponding to the phase
of quasar activity.  We assume the timescale for exponential decay is
$\sim1$\,Gyr, similar to that estimated in simulations neglecting
black hole feedback, and demand that the stellar mass after multiple
$e$-foldings is that given by e.g.\ our $\mbh-\mbul$ relation
(although this choice only weakly effects our results, so long as the
$\mbh-\mbul$ relation holds at least approximately after $\sim1$ or
more $e$-foldings in the star formation rate). The primary result of
this is indicated in the lower right panel of the figure, namely that
the galaxies are much too blue (by $\sim1$ magnitude), and do not
develop the characteristic slope of the color-magnitude relation.
This demonstrates the dramatic importance of black hole feedback, as
the rapid quenching of star formation both allows remnants to redden
sufficiently and enables the gradient in formation age with mass to
produce a slope in the color-magnitude relation, as opposed to its
being ``washed-out'' by continued star formation in hosts of all
masses, regardless of the peak in their star formation histories.

\begin{figure}
    \centering
    \epsscale{1.15}
    \plotone{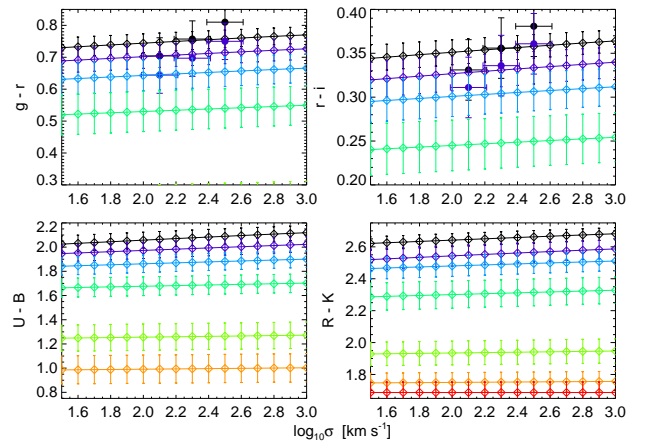}
    \caption{Predicted mean color (diamonds and lines) as a function
    of spheroid stellar velocity dispersion, in the manner of
    Figure~\ref{fig:color.mag}. Again, our predictions are shown for
    $z=0$ (black), $z=0.2$ (purple), $z=0.5$ (blue), $z=1$ (cyan),
    $z=2$ (green), $z=3$ (orange), $z=5$ (red), with bluer colors at
    higher redshift. At $z=0$ and $z=0.2$, our predicted $(g-r)$ and
    $(r-i)$ colors are compared to those observed as a function of
    velocity dispersion in \citet{BernardiIV,Bernardi05}.
    \label{fig:color.sigma}}
\end{figure}

Figure~\ref{fig:color.sigma} shows the predicted colors of remnant
spheroids as a function of spheroid stellar velocity dispersion and
redshift (assuming pure density evolution above $z\sim2$).  We
consider the colors SDSS $(g-r)$ (upper left) and $(r-i)$ (upper
right) and the standard $(U-B)$ (lower left) and $(R-K)$ (lower right)
colors. For the $(g-r)$ and $(r-i)$ colors, we compare to the
color-$\sigma$ relations observed by \citet{BernardiIV,Bernardi05}
(filled circles) at $z=0$ (black) and $z=0.2$ (purple).  Both the
$z=0$ mean colors and their evolution at low redshift are reproduced
by our modeling, but this is not trivial even given the $M_{\rm
BH}-\sigma$ relation and fundamental plane, as for example the scatter
in color is not equivalent as a function of luminosity or velocity
dispersion.  The dependence on velocity dispersion is also reasonably
well described, with our prediction within $1\sigma$ of the
observations over the range of velocity dispersion for which they
exist.

\begin{figure}
    \centering
    \epsscale{1.15}
    \plotone{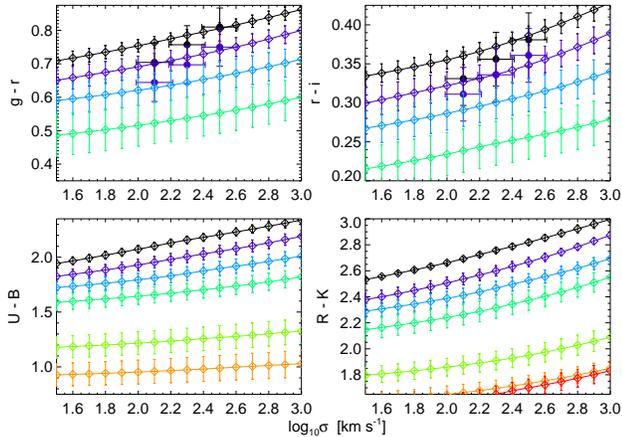}
    \caption{Same as Figure~\ref{fig:color.sigma}, but adopting the maximal 
    dependence of total metallicity on age and velocity dispersion from \citep{Jorgensen99b}.
    \label{fig:color.sigma.metal}}
\end{figure}

The weak variation in these colors with velocity dispersion, however,
means that the small effects of a systematic dependence of total
metallicity on velocity dispersion or age may be important. We show
the consequences of such a dependence in
Figure~\ref{fig:color.sigma.metal}, where we repeat the modeling of
Figure~\ref{fig:color.sigma}, but adopt a scaling of metallicity with
age (here we mean $z=0$ age, i.e.\ formation redshift) and velocity
dispersion.  To estimate the maximum effect, we consider a metallicity
dependence following the strongest scaling of [Fe/H] with age and
velocity dispersion found by \citet{Jorgensen99b}, namely $[{\rm
Fe/H}]=-0.46\, \log({\rm age/Gyr})+0.33\,\log(\sigma/{\rm
km\,s^{-1}})-0.30$.  We choose this scaling as opposed to others
\citep[e.g.,][]{Kuntschner00} because it includes both the variation
with age and velocity dispersion, but we find similar results
neglecting the dependence on age.  The resulting color-magnitude
relations are steepened, and their slopes agree well with the
observations.  The colors change by a negligible amount at the
approximate zero-point of the observations at $\sigma\sim200\,{\rm
km\,s^{-1}}$, because here the offset of the color-magnitude relation
is determined by the ages of the spheroid populations alone, and
agrees well as in Figure~\ref{fig:color.sigma}.  Also, although the
agreement in slope appears improved, we note that the effect is still
small, generally $\lessim0.05$\,mag in a given color even at the
extreme values of $\sigma\approx30,\ 1000\,{\rm km\,s^{-1}}$ plotted
(except for the high-$\sigma$ end of the $(R-K)$ colors, which are
discussed above in greater detail).

This is an approximate upper limit, for example the other
determinations within \citet{Jorgensen99b} yield smaller logarithmic
slopes of metallicity with $\sigma$, e.g.\ $\sim0.07$ as opposed to
the $0.33$ shown.  That this is a still small effect and further that
it serves to bring our predictions into better agreement with
observations, suggests that we are safe in neglecting it in other
predictions.  However, with improved observations of the
color-$\sigma$ variation, the distinctions between the predictions in
e.g.\ Figure~\ref{fig:color.sigma} and
Figure~\ref{fig:color.sigma.metal} could be significant enough to
constrain the strength of the metallicity evolution allowed or
required.

We find that the scatter in colors at a given $\sigma$ is typically
smaller than that at a given magnitude. In \S~\ref{sec:ages} below, we
demonstrate that this is a consequence of the fact that velocity
dispersion is directly related to the black hole masses forming over
cosmic time, whereas the $z=0$ magnitude mixes systems of different
masses and ages (and thus different colors) at the same observed
luminosity.  Observationally, \citet{BernardiIV,Bernardi05} also find
that these correlations have small scatter, similar to our
predictions, and argue that they are tighter and may represent a more
fundamental correlation than, e.g.\ the color-magnitude relations.  We
also note that the qualitative behavior of colors as a function of
velocity dispersion and redshift is similar for each of the colors
considered, although different colors are rescaled about different
values, and the evolution in the slope of the color-$\sigma$ relation
is much weaker than that of the color-magnitude relation.  These
properties make the color-velocity dispersion relation a valuable
probe not just as a check on the color-magnitude relation but
potentially as a measurement independent of some systematics (for
example, the common observational 
assumption of constant slope with redshift in this case
appears quite reliable).

\begin{figure}
    \centering
    \epsscale{1.15}
    \plotone{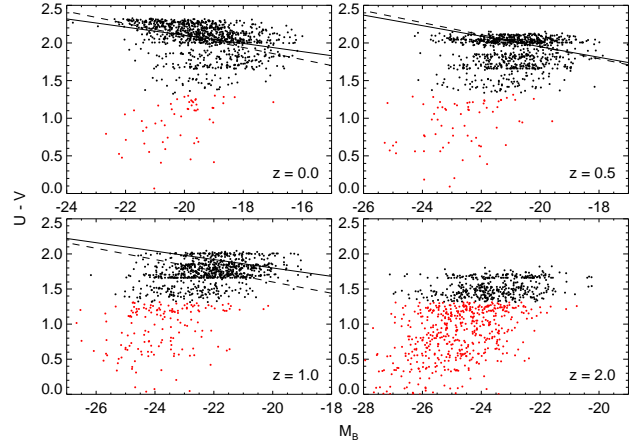}
    \caption{Predicted $(U-V)$ vs.\ $M_{B}$ color-magnitude relation at redshifts $z=0,\ 0.5,\ 
    1,\ {\rm and}\ 2$, as labeled. In each panel, 1000 galaxies are generated according to the 
    predicted joint color-magnitude distributions at the given redshift. Black points show 
    galaxies older than 0.5\,Gyr, red points younger. In the upper left, the solid line shows 
    the best-fit color-magnitude to our predictions, the dashed line the best-fit to the 
    observed galaxies from \citet{Bell04b,Giallongo05}. In the upper right and lower left, the solid line 
    shows the observed color-magnitude relation of \citet{Giallongo05}, dashed line the observed 
    relation of \citet{Bell04b}.
    \label{fig:mock.color.mag}}
\end{figure}

Finally, we use our modeling to generate an observed color-magnitude
relation in Figure~\ref{fig:mock.color.mag}.  At each redshift
considered, we calculate the joint probability distribution in both
color and magnitude based on our predicted history of spheroid
formation prior to that redshift (i.e.\ the color distribution at a
given magnitude in Figure~\ref{fig:color.mag}, and distribution in
magnitudes from our predicted luminosity functions in e.g.\
Figure~\ref{fig:LF.B}), and generate 1000 points (mock galaxies)
according to that probability distribution.  These are not full
simulated galaxies, but random points drawn from our calculated
joint PDF in color and magnitude at each redshift.  At $z=0$, we
directly fit the generated points to a color-magnitude relation, and
show the result, $(U-V)=1.9-0.04\,(M_{B}+20)$ as a solid black
line. Our result is similar to the observed relation,
$(U-V)=2.1-0.08\,(M_{B}+20)$ from \citet{Bell04b} and
\citet{Giallongo05}, as is the absolute distribution in magnitude and
color.  We show galaxies older than 0.5\,Gyr as black points, and
galaxies younger than this as red points. This demonstrates that very
young galaxies are not a significant contributor to the observed red
galaxy population at low redshift, and thus the fact that they lie in
a more blue, brighter region of color-magnitude space than the
``normal'' relaxed elliptical population, as well as most likely being
disturbed systems which would not be morphologically recognized as
ellipticals, is not important in our calculations at low redshift. The
removal of these points at $z=0$ does not change our results
significantly, except to slightly steepen the fitted color-magnitude
slope to $-0.06$, in better agreement with that observed.

At $z=0.5$ and $z=1$, the fractional ``young'' population is still
relatively small, although it does increase, and the observed
color-magnitude relations still agree well with our predicted
distribution of ``old'' elliptical colors and magnitudes.  We show the
observed color-magnitude relations of \citet{Bell04b}, who assume a
constant slope at all redshifts, at these redshifts as dashed lines,
and the observed relation of \citet{Giallongo05}, who allow the slope
to vary, as solid lines.  As shown in Figure~\ref{fig:color.mag.old},
we reproduce the observed evolution in the red/elliptical
color-magnitude relations if we restrict ourselves to the older
spheroids which have had sufficient time after their progenitor
gas-rich mergers to relax and be recognized as ellipticals by either
color or morphological selection criteria. By $z=2$ (lower right),
however, the fraction of young objects becomes quite large
($\approx0.5$), as observed and discussed further above and in
\S~\ref{sec:ages}.

\section{Spheroid Mass to Light Ratios and Luminosity-Size Relations as a Function of Mass and Redshift}
\label{sec:ML}

\begin{figure*}
    \centering
    \plotone{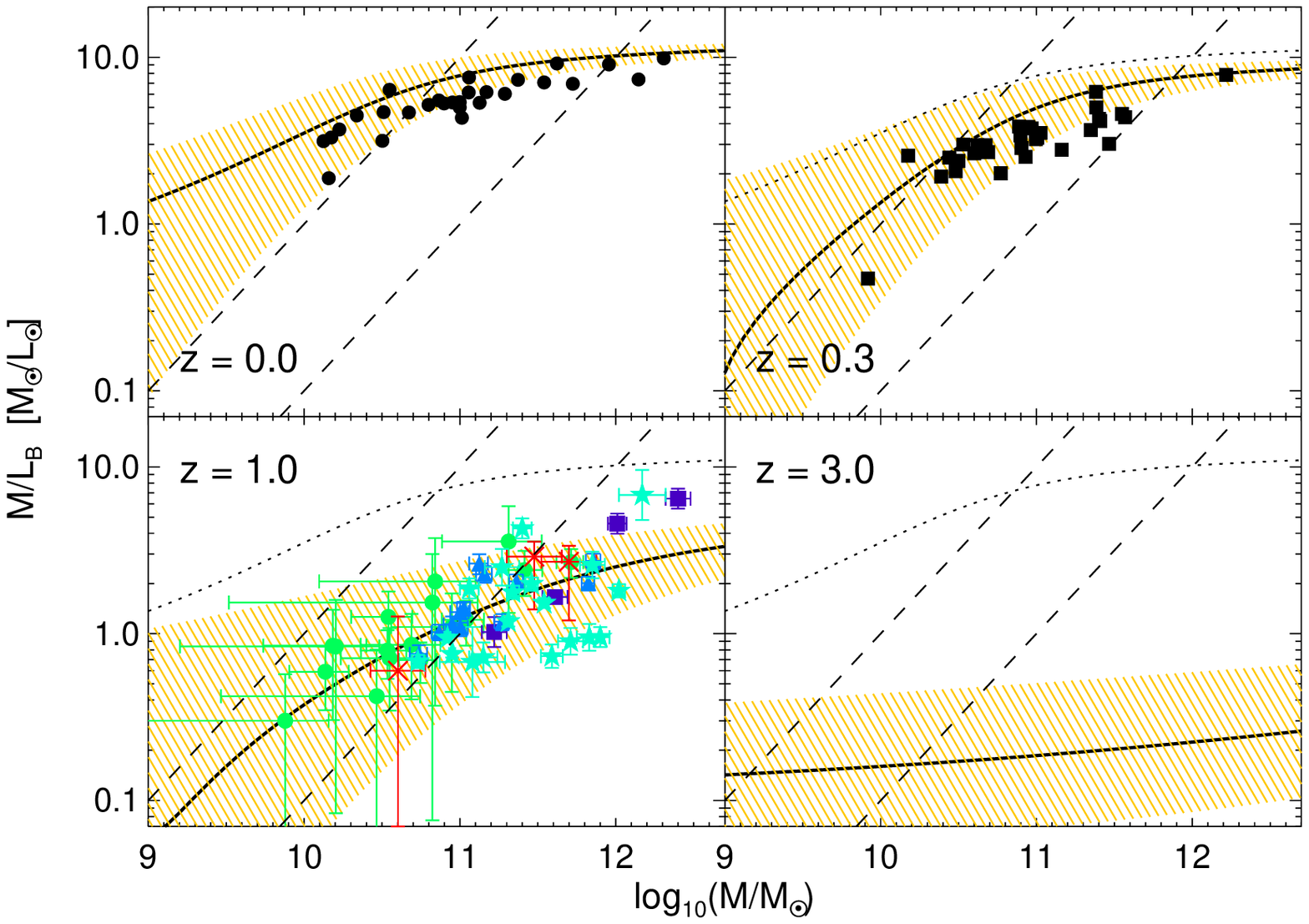}
    \caption{Predicted mean mass to B-band light ratio $M/L_{B}$ (solid lines) 
    as a function of spheroid mass $M$, with $1\sigma$ dispersion at each $M$ (yellow shaded region). 
    Results are shown for $z=0$ (upper left), $z=0.3$ (upper right), $z=1$ (lower left), and $z=3$ (lower right), 
    as labeled. Observations at $z\approx0$ (black circles) are from 
    \citet{JFK95a,JFK95b,Jorgensen96}, at $z\approx0.3$ (black squares) from \citet{Kelson00}, 
    and at $z\approx1$ from 
    \citet{vdW05} (cyan stars), \citet{Holden05} (purple squares), \citet{vDS03} (red $\times$'s), 
    \citet{Wuyts04} (blue triangles), and \citet{SA05} (green circles). 
    Luminosity limits of $10^{10}\,L_{\sun}$ and $10^{11}\,L_{\sun}$ are shown in 
    each panel (dashed lines), as is the $z=0$ mean $M/L_{B}$ (dotted lines). 
    \label{fig:ML}}
\end{figure*}
\begin{figure}
    \centering
    \epsscale{1.15}
    \plotone{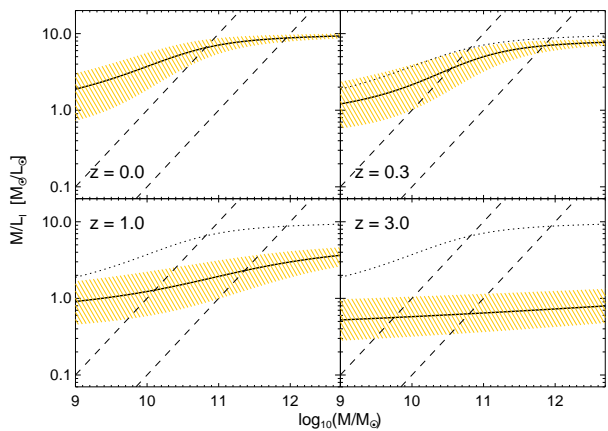}
    \caption{As Figure~\ref{fig:ML}, but our predictions are shown for the 
    mass to I-band mass to light ratio $M/L_{I}$.
    \label{fig:ML.i}}
\end{figure}

Figure~\ref{fig:ML} shows our predicted $M/L$ ratio in the B-band
($M/L_{B}$) as a function of spheroid mass.  For each redshift, we use
our modeling of $\nBH$, $\nbul$ from the quasar luminosity function to
determine the distribution of ages for spheroids of a given mass at
that redshift, and from that determine the distribution of $M/L$
ratios in a given band.  The masses shown are $\mvir$, the virial mass
within the effective radius ($\equiv5\sigma^{2}R_{e}/G$, as defined in
\S~\ref{sec:scaling}), in order to ease comparison with
observations (which generally adopt this choice; those that do not
have been rescaled accordingly).

Our $z=0$ prediction is compared to observations of spheroids in the
Coma cluster (at $z=0.023$) from \citet{JFK95a,JFK95b,Jorgensen96}
(black circles), which are similar to recent determinations from the
SDSS and other studies \citep[e.g.][]{vdW05,Cappellari05}. The $z=0.3$
result is compared to observations of the cluster Cl 1358+62 at
$z=0.33$ from \citet{Kelson00} (black squares). Our $z=1$ prediction
is compared to several different observations, including those from
$0.6<z<1.15$ in the Chandra Deep Field-South sample of \citet{vdW05}
(cyan stars), the $z=1.237$ cluster RDCS 1252.9-2927 sample of
\citet{Holden05} (purple squares), the $z=1.27$ cluster RDCS
J0848+4453 galaxies from \citet{vDS03} (red $\times$'s), the $z=0.83$
cluster MS 1054-03 sample of \citet{Wuyts04} (blue triangles), and the
$0.88<z<1.3$ K20 sample of \citet{SA05} (green circles). In each panel
at $z>0$, we show the $z=0$ mean $M/L_{B}$ prediction for comparison
(dotted lines). We also show our predicted mass to I-band light ratios
$M/L_{I}$ as a function of mass in Figure~\ref{fig:ML.i}, in the same
manner as Figure~\ref{fig:ML}, demonstrating the relative importance
of different age distributions in different observed wavebands.

Our modeling reproduces the typical $M/L_{B}$ ratios and their
dependence on mass, and the scatter about the mean $M/L_{B}$, which
increases significantly with increasing redshift and decreasing
mass. Although for clarity we have not shown other redshifts, we have
compared e.g.\ the $z=0.58$ MS 2053-04 sample of \citet{Wuyts04} to
our predictions and find similar agreement. Our modeling further
predicts the observed {\em differential} evolution in $M/L_{B}$, where
the mass to light ratio declines more rapidly with redshift above
$z=0$ in smaller-mass systems, implying that these formed more
recently \citep[see, e.g.][]{Treu01,vanDokkum01,Treu02,
vDS03,Gebhardt03,Rusin03,vandeVen03,Wuyts04,Treu05,Holden05,vdW05,SA05}.
At $z\gtrsim2$, our model agrees well with the observations, for
example the mass-to-light ratio as a function of mass in the K-band of
distant red galaxies found by \citet{Labbe05}, which may even observe
the flattening in the $M/L$ relation we predict for $z\gtrsim2-3$,
although it is difficult to determine this given luminosity limits at
these high redshifts.  These observations suggest that many of the
most massive galaxies are forming at this redshift, with $\sim70\%$ of
the population being blue, dusty galaxies still forming stars at a
high rate \citep{Labbe05}, as we expect (see \S~\ref{sec:ages} for a
more detailed discussion) and a fraction of the most massive galaxies
formed as early as $z\sim5$, although this age is lower than estimated
in e.g.\ Labb{\'e} et al.\ (2005) if we account for the rapid
quenching of star formation seen in our simulations in modeling the stellar populations
\citep[e.g.][]{FS04,vanDokkum04}.

Our modeling suggests that the $M/L_{B}$ relation should steepen below
$M\sim{\rm a\ few}\ \times 10^{10}\,M_{\sun}$, where at low redshift,
samples are severely limited by luminosity/magnitude limits, making
the differential evolution slightly less dramatic. However, we caution
against interpreting this curvature too strictly, as it depends on
both the functional form and quantitative dependence of the quasar
luminosity function break luminosity on redshift.  In our adopted form
for the quasar luminosity function, the break luminosity evolves
exponentially with lookback time, in which case the degree of
curvature is quite sensitive to the coefficient of this exponential
growth, whereas if e.g.\ we considered exponential evolution in
redshift (instead of lookback time), we obtain similar values of
$M/L_{B}$ at small and large $M$, but with a less curved power-law
interpolation between them.

To illustrate the impact of selection effects, we plot (dashed lines)
the lower observable mass limit for a limiting luminosity of
$10^{10}\,L_{\sun}$ (left) and $10^{11}\,L_{\sun}$ (right) in each
panel.  The scaling we describe in \S~\ref{sec:scaling} between virial
and stellar mass within the effective radius (or stellar mass and
effective radius) is a non-negligible component of the $z=0$ slope of
the $M/L$ ratio -- ignoring this scaling does not change our
predictions at the high-mass end, but results in an overprediction of
the $M/L$ ratio at the low-mass end by a factor $\sim2$.  However, the
redshift evolution is almost entirely a consequence of the different
ages of spheroids of different mass; our predictions for the
differential $M/L$ evolution with redshift are essentially identical
if we neglect the weak evolution in the $\mbul-R_{e}$ relation with
redshift described in Robertson et al.\ (2005c, in preparation).

Differential evolution in the $M/L_{B}$ ratio is expected in our model
because the break in the quasar luminosity function shifts to lower
luminosities below $z\sim2-3$, implying that spheroids with smaller black
hole mass (smaller peak luminosity) are dominating the distribution of
objects being formed at these later times. Therefore, at $z\sim1$, the
lower mass objects have formed more recently.  However, above
$z\sim2-3$, this differential evolution should either flatten or
reverse, if a pure density or pure peak luminosity evolution
model of the quasar luminosity function is an accurate description of
quasar activity. The results in Figure~\ref{fig:ML} assume pure
density evolution in the quasar luminosity function above $z\sim2$.
In this case, above $z\sim2$, the shape of the luminosity function
(and therefore, the distribution of peak luminosities and
corresponding spheroid masses being formed) remains constant, and the
normalization decreases with higher redshift. Thus, all objects have
the {\em same} distribution of formation ages above this redshift
(with only second-order effects from the finite quasar lifetime and
merger time, at least until high redshifts where these times become
comparable to the Hubble time). Therefore, the slope of $M/L_{B}$ vs.\
$M$ should become flat (except for the small effects of the
$\mbul-R_{e}$ relation), as seen in the figure for $z=3$.

In a pure peak luminosity evolution scenario, the shape of the quasar
luminosity function above $z\sim2$ again remains roughly constant, but
instead of decreasing in normalization, the break luminosity shifts to
smaller luminosities at higher redshifts, with constant
normalization. This implies that, above $z\sim2-3$, the more massive
objects have actually formed more recently, and so the slope of the
$M/L_{B}$ vs.\ $M$ relation should be inverted, i.e.\ that $M/L_{B}$
should decrease with mass.  However, if metallicity evolves with
either mass or redshift, this will affect the mean mass to light ratio
and slope as well, although we discuss this effect above and show in
Figure~\ref{fig:color.sigma.metal} that it is small.

\begin{figure}
    \centering
    \epsscale{1.15}
    \plotone{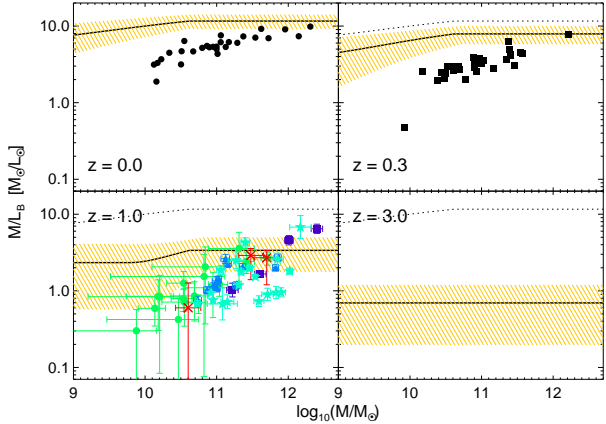}
    \caption{As Figure~\ref{fig:ML}, but assuming a ``light-bulb'' (``on/off'') 
    or pure exponential model of the quasar light curve and lifetime. 
    \label{fig:ML.simplelife}}
\end{figure}

We also test whether the distributions of spheroid mass to light ratios inferred
from idealized models of the quasar lifetime are consistent with
observations.  We consider a case in which quasars have a fixed,
constant lifetime and radiate at a fixed luminosity $L=\Lp$.  Here,
the value of the quasar lifetime is unimportant, as it controls only
the normalization of the resulting rates of spheroid formation.  We
adopt the luminosity function of \citet{Ueda03}, from the hard X-ray,
modified for pure density evolution above $z=2$ following
\citet{Fan01}, although our results are qualitatively insensitive to
these specific choices \citep{H05c}.  We have already demonstrated in
Figure~\ref{fig:LF.B} and Figure~\ref{fig:color.mag.tracks} that such
modeling predicts a spheroid luminosity function and color-magnitude
relation in stark disagreement with observations.

Figure~\ref{fig:ML.simplelife} shows the predicted B-band mass to
light ratios $M/L_{B}$ as a function of mass at redshifts $z=0,\ 0.3,\
1,\ 3$, in the same manner as Figure~\ref{fig:ML} and with the same
observations shown, but adopting this idealized model for the quasar
light curve. The predicted mass-to-light ratio is too high by a factor
$\sim2-5$ at all but the largest masses, and shows almost no
dependence on mass at any redshift, and no differential evolution
from $z=0$ to $z=1$.  Although both the color-magnitude relation
and mass to light ratios derive from the same underlying age
distribution, the distinction between the predictions of our full
model of quasar activity and idealized models is significantly 
stronger in the predicted mass to light ratios than color magnitude
relations (Figure~\ref{fig:color.mag.tracks}).  We note that the
\citet{Ueda03} luminosity function does include ``luminosity-dependent
density evolution,'' in which the slope of the faint-end quasar
luminosity function evolves with redshift, implying that the density
of lower-luminosities quasars peaks at lower redshift. This is the
only reason, in fact, that there is any dependence of $M/L_{B}$ on $M$
at all in Figure~\ref{fig:ML.simplelife}.  Although this is
qualitatively consistent with the anti-hierarchical, downsizing
picture implied by the observations described above, the figure
demonstrates that it is quantitatively {\em insufficient} to account
for the downsizing observed in the spheroid population.  

At high
redshifts $z\sim2$, traditional models of the quasar luminosity
function associate an observed luminosity with a quasar's {\em peak}
luminosity, implying that many low-peak luminosity (i.e.\ low final
black hole mass and, correspondingly, small spheroid mass) systems are
forming at these redshifts.  Even if the inferred formation of these
objects reaches a maximum at somewhat lower redshift, they are still
formed over a wide range of redshifts with a large number of the
smallest-mass systems formed at $z\sim1-3$. However, in our model
these observed faint-end objects are really brighter peak luminosity
sources, in a dimmer stage of their evolution; the distribution of
peak luminosities being formed at a given redshift is actually {\em
peaked}, at a luminosity corresponding the break in the observed
luminosity function. Thus, low peak luminosity systems (small spheroid
masses) are not formed until much later times, when the {\em break}
luminosity has evolved to small luminosities. In fact, in our
modeling, the observed change in quasar luminosity function slope is
actually a consequence of the quasar lifetime as a function of
luminosity, while the break luminosity evolution reflects ``cosmic
downsizing'' \citep{H05f}.

\begin{figure}
    \centering
    \epsscale{1.15}
    \plotone{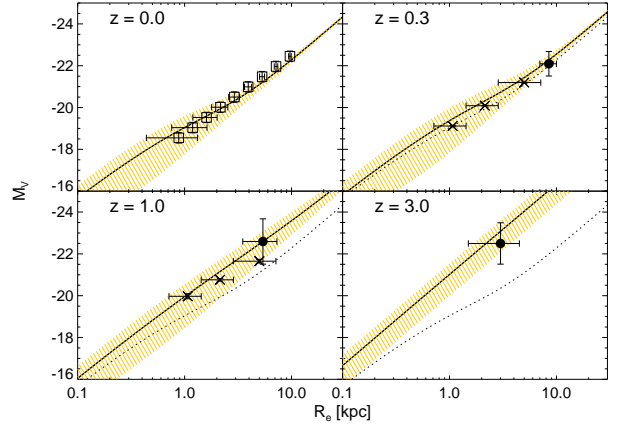}
    \caption{Predicted luminosity-size relation (in V-band) at several redshifts, as labeled. 
    The mean luminosity-size relation (black lines) and $1\sigma$ range (yellow shaded area) are shown. 
    Dotted lines in each panel show the mean $z=0$ relation. 
    Observations at $z=0$ (squares) are from \citet{Shen03}, with horizontal error bars showing 
    the dispersion in $R_{e}$ at each constant $M_{V}$. Observations at $z>0$ are from 
    \citet[][circles]{Trujillo05} and \citet[][$\times$'s]{McIntosh05b}.
    \label{fig:LR}}
\end{figure}
\begin{figure*}
    \centering
    \plotone{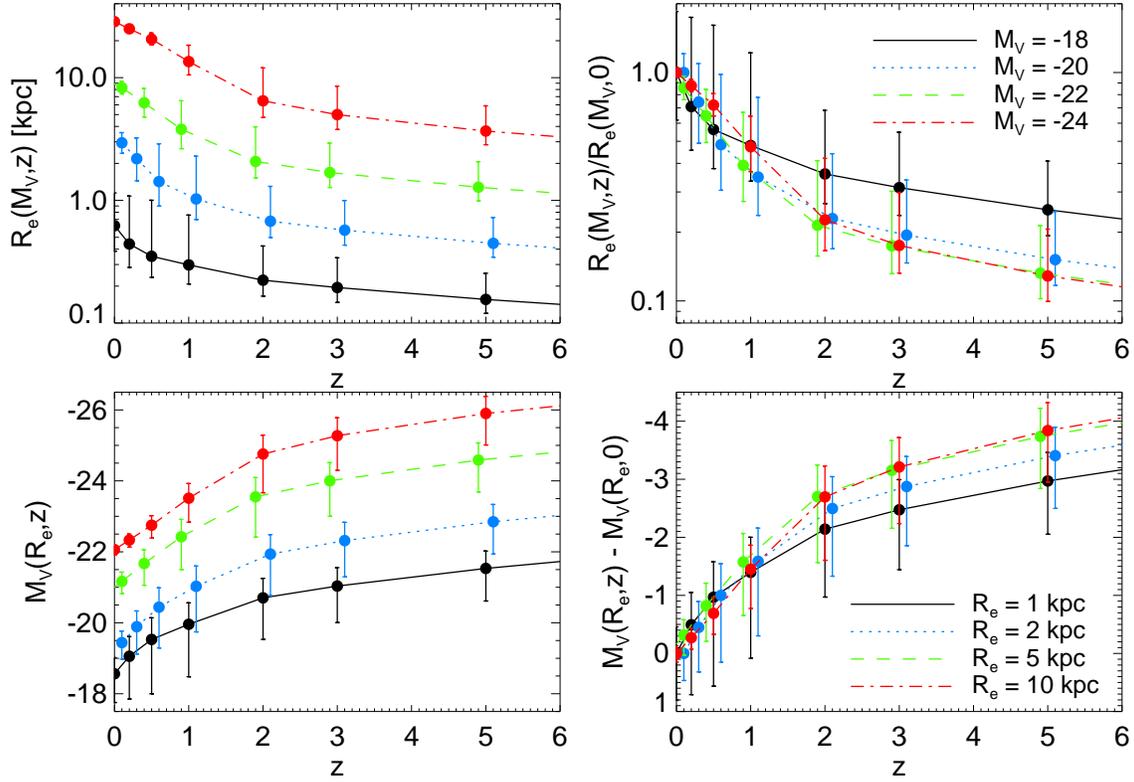}
    \caption{Predicted luminosity-size relation (in V-band) as a function of redshift. 
    Upper panels show the absolute (left) and relative (normalized to the $z=0$ value, right) 
    effective radii $R_{e}$ (and $1\sigma$ range of radii, vertical error bars) as a function 
    of redshift, at fixed luminosity ($M_{V}=-18,\ -20,\ -22,\ -24$, as labeled).  
    Lower panels show the absolute (left) and relative (right) 
    V-band magnitude (and $1\sigma$ range of magnitudes, vertical error bars) as a function 
    of redshift, at fixed effective radius ($R_{e}=1,\ 2,\ 5,\ 10$\,kpc, as labeled).  
    \label{fig:LR.z}}
\end{figure*}

As discussed in \S~\ref{sec:scaling} above, Robertson et al.\ (2005c,
in preparation) analyze scaling relations for merger remnants and
their implications for the fundamental plane.  However, that work
considers only the structural properties of individual objects and does
not predict the age distribution of any population.  Here, we
determine the distribution of spheroid ages as a function of e.g.\
stellar mass, and combine this with knowledge of the detailed
structure of the remnants to predict the observed luminosity-size
relations as a function of redshift in bands where mass to light
evolution is important. For present purposes, we emphasize that our
simulations reproduce well the observed $z=0$ effective radius-stellar
mass relation of remnant red/elliptical galaxies (e.g. Bernardi et
al.\ 2003a, Shen et al.\ 2003; Padmanabhan et al.  2004; Cappellari et
al. 2005), as well as predicting that this relation should evolve at
most weakly with redshift, in agreement with observations
\citep{Trujillo04a,Trujillo04b,Trujillo05,McIntosh05b} \citep[see also
e.g.][although these authors do not separate the relation by
morphological type] {Ferguson04,Bouwens04,Papovich05}.  Given a nearly
redshift-independent $R_{e}-\mbul$ relation, it is then
straightforward to convert our predicted mass-to-light ratios as a
function of mass to a luminosity-size relation (luminosity as a
function of effective radius). This then enables a secondary means of
measuring the relative ages and differential evolution of the remnant
spheroid population, which in many cases probes different regimes in
size and redshift.

Figure~\ref{fig:LR} shows the resulting predicted luminosity-size
relation (in V-band) at several redshifts.  We compare to observations
at $z=0$ (squares) from \citet{Shen03}, with horizontal error bars
showing the dispersion in $R_{e}$ at each constant $M_{V}$. These
observations are converted from the r-band using our predicted
color-magnitude relations (\S~\ref{sec:colors}), which further
implicitly guarantee that we reproduce the observed luminosity-size
relation in all other wavebands.  Further observations at each $z>0$
are shown from \citet[][circles]{Trujillo05} and
\citet[][$\times$'s]{McIntosh05b}.

Our modeling reproduces both the mean luminosity-size relation at each
redshift, as well as the range of $R_{e}$ at fixed luminosity as a
function of luminosity (compare the $z=0$ dispersions from Shen et
al.\ (2003) and our modeling).  For the observed redshift ranges, the
effect of the change in the $\mbul-R_{e}$ relation with redshift in
our simulations is small, for example at fixed
$\mbul=10^{10}\,M_{\sun}$ the effective radius decreases by just
$25\%$ from $z=0$ to $z=2$, and the evolution in the luminosity-size
relation is driven primarily by evolution in mass-to-light ratios
owing to different spheroid ages as a function of mass or size.

We show the evolution with redshift of both effective radius at fixed
luminosity and luminosity at fixed effective radius in greater detail
in Figure~\ref{fig:LR.z}.  The points at each magnitude are offset by
a negligible amount for clarity.  Although the interpretation is not
as straightforward as that of our mass-to-light ratio predictions, the
more rapid and pronounced relative magnitude evolution of systems with
larger effective radii is a reflection of the same anti-hierarchical
growth discussed above (and below in \S~\ref{sec:ages}), with larger
(higher-mass) systems forming at higher redshift.

\section{Galaxy Ages as a Function of Mass and Luminosity}
\label{sec:ages}

\begin{figure*}
    \centering
    \epsscale{1.1}
    \plotone{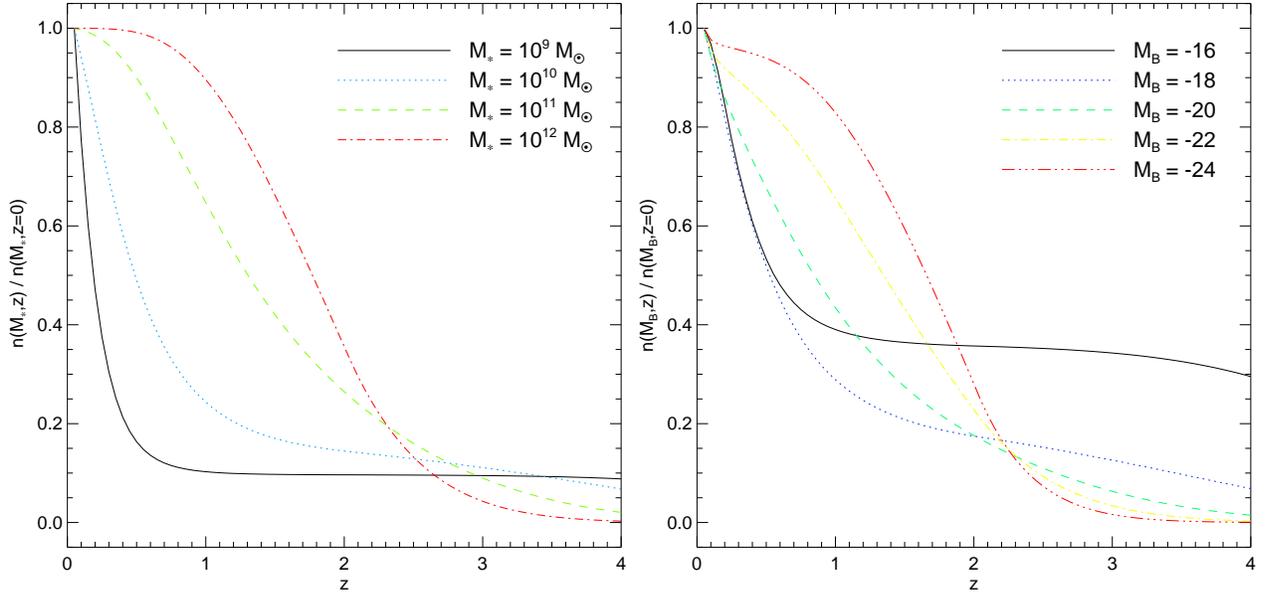}
    \caption{Predicted fraction of $z=0$ spheroids with stellar mass
    $M_{\ast}$ formed by a given redshift as a function of redshift,
    for $M_{\ast}=10^{9},\ 10^{10},\ 10^{11},\ {\rm and}\
    10^{12}\,M_{\sun}$, as labeled (left). Right panel shows the same,
    but for spheroids observed at $z=0$ with a given B-band magnitude
    $M_{B}=-16,\ -18,\ -20,\ -22,\ -24$, as labeled.
    \label{fig:ages}}
\end{figure*}

Figure~\ref{fig:ages} shows the fraction of all $z=0$ spheroids of a
given stellar mass formed by a given redshift, as a function of
redshift for spheroid stellar masses $M_{\ast}=\mbul=10^{9},\
10^{10},\ 10^{11},\ {\rm and}\ 10^{12}\,M_{\sun}$.  Given the
anti-hierarchical nature of black hole growth described in
\citet{H05e}, where the highest-mass black holes are formed at high
redshifts, associated with the peak in bright quasar activity, and
lower mass black holes are formed at lower redshift as the break in
the observed quasar luminosity function (corresponding to the peak in
the formation rate of final black hole masses $\nBH$) moves to lower
luminosities, we expect the trend indicated, where higher-mass
spheroids are formed at higher redshifts and over a wider range in
redshift, as these correspond to higher-mass black holes.

This evolution in black hole mass explains the observations of
\citet{BernardiIV,Bernardi05}, who find that color is primarily
correlated with velocity dispersion (see
Figure~\ref{fig:color.sigma}), with the color-magnitude relations
discussed above being a consequence of the fact that magnitudes are
also correlated with velocity dispersion. Based on the quasar
luminosity function, the dispersion in ages for a given $\sigma$ is
small, as black holes of a given mass form over a well-defined range
of redshifts. Since feedback from black hole growth results in passive
evolution of the remnant after quasar activity, the age (and therefore
reddening) of the remnant is correlated more tightly with the velocity
dispersion (i.e.\ black hole mass) of the remnant than its luminosity
(magnitude), which mixes galaxies of different black hole masses and
ages.

Figure~\ref{fig:ages} also shows the fraction of all $z=0$ spheroids
of a given B-band magnitude formed by a given redshift.  Unlike the
fractional population vs.\ redshift as a function of mass, this
includes the effects of stellar evolution, effectively mixing e.g.\
older, more massive galaxies with younger, less massive ones that have
the same $z=0$ B-band luminosity.  Despite this, the trend of higher
luminosity objects forming at characteristically larger redshifts and
over a wider range of redshifts is clear.  The flattening of the
lowest-luminosity population growth at $z\sim1$ is a consequence of
the pure peak luminosity evolution model for the quasar luminosity
function evolution at $z\gtrsim2$. With pure density evolution above
$z\sim2$, the lowest luminosity curve will continue to fall rapidly,
without a significant number of very low peak luminosity (low spheroid
mass) systems forming at high redshift.

\begin{figure}
    \centering
    \epsscale{1.15}
    \plotone{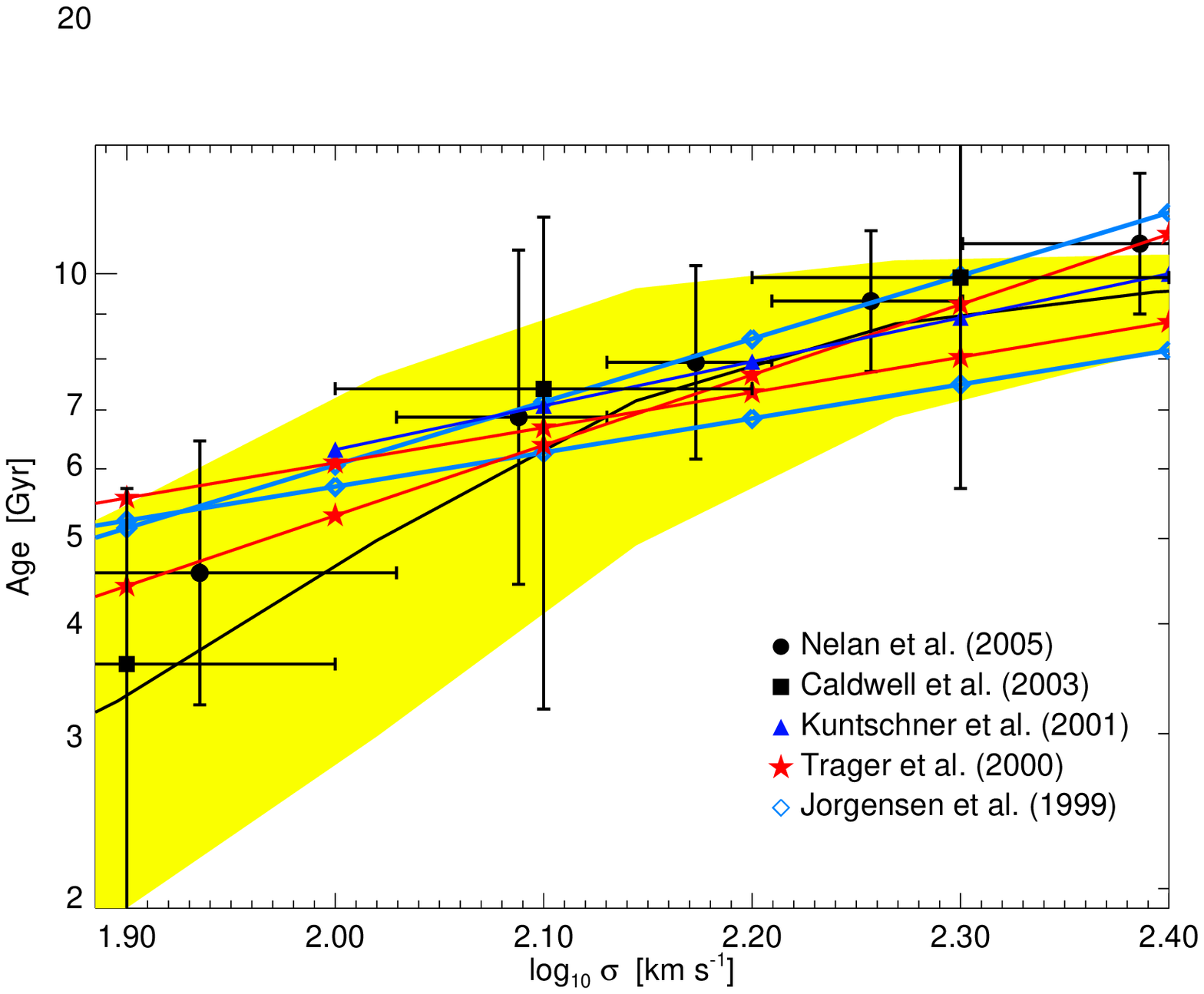}
    \caption{Predicted $z=0$ ages of spheroids as a function of 
    velocity dispersion. Black solid line shows our predicted median age, 
    yellow range shows the interquartile range of ages at each $\sigma$. 
    Observations of mean age (and dispersion about the mean, vertical 
    error bars) in bins of $\log\sigma$ are shown from 
    \citet{Nelan05} (circles) 
    and \citet{Caldwell03} (squares), and the $\sim1\sigma$ range of fitted 
    age-$\sigma$ relations from \citet{Kuntschner01} (dark blue line with triangles), 
    \citet{Trager00} (red line with stars), and \citet{Jorgensen99} (light blue line with diamonds) 
    are shown as solid lines. 
    \label{fig:age.vs.sigma}}
\end{figure}

In Figure~\ref{fig:age.vs.sigma}, we plot the predicted $z=0$ ages of
spheroids as a function of velocity dispersion from our modeling
(assuming pure peak luminosity evolution at $z\gtrsim2$, although this
only becomes important here at very low $\sigma$ where the range of
ages is relatively large in either case). 
Observations of the mean age in bins of $\log\sigma$ are shown from
\citet{Nelan05} (circles) and \citet{Caldwell03} (squares), with
horizontal errors showing the range of $\log\sigma$ of each bin and
vertical error bars showing the rms dispersion in ages at the given
velocity dispersion (which can be compared to the yellow range
plotted). The $\sim1\sigma$ range of fitted age-$\sigma$ relations
(i.e.\ adopting the minimal and maximal fitted age-$\sigma$ slopes)
from \citet{Kuntschner01} (dark blue line with triangles),
\citet{Trager00} (red line with stars), and \citet{Jorgensen99} (light
blue line with diamonds) are shown as solid lines. The slopes from the
observations of \citet{Kuntschner01} and \citet{Trager00} are
determined by fitting in \citet{Nelan05}.  

The agreement at all values of $\sigma$ is good, again implying that
the downsizing of both galaxy and quasar populations is
self-consistent when our model of the quasar lifetime is adopted, and
emphasizing that age evolution as a function of velocity dispersion or
stellar mass is important along the red sequence (i.e.\ that the red
sequence is not merely a metallicity sequence).  There is a slight
systematic offset in the mean age, with several of the observations
estimating ages $\sim1\,$Gyr larger than those we predict, but this is
well within the uncertainties of both our theoretical modeling and 
observational estimates of absolute ages.  

Our prediction of the age-velocity dispersion relation includes the
observed steepening of the relation at low velocity dispersions
\citep[e.g.,][]{Caldwell03,Nelan05}, an
effect not accounted for in fitting a single power law, which is why
the power law fits extrapolated to low $\sigma$ tend to predict larger
ages than given by either our prediction or the binned
observations. There is also a suggestion that the dispersion in age
becomes larger at low velocity dispersion, an effect discussed in
detail in \S~\ref{sec:colors}, and potentially seen in some
observations \citep[e.g.][]{Nelan05}, but the observations are still
uncertain on this point and as shown in regard to the color-magnitude
relation, this effect can be quite sensitive to whether pure peak
luminosity or pure density evolution is assumed for the quasar
population at high redshift.
 
\begin{figure*}
    \centering
    \plotone{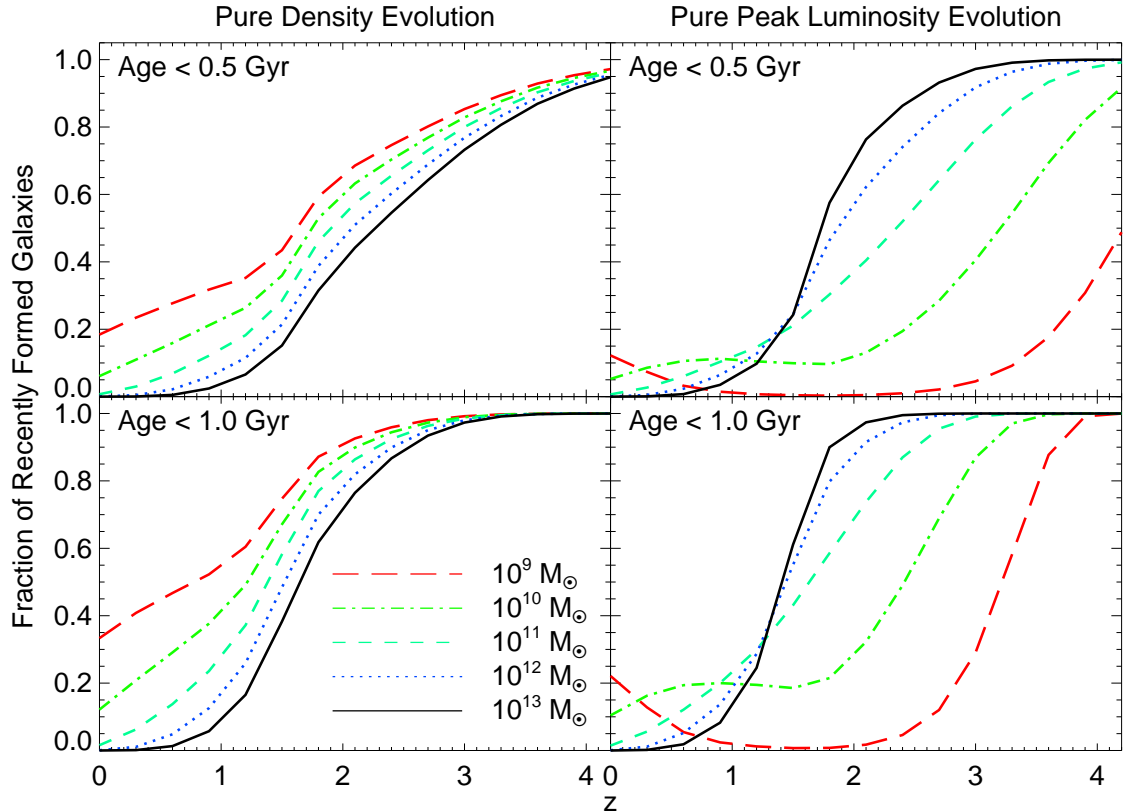}
    \caption{Predicted fraction of spheroids of a given total 
    stellar mass (as labeled) with ages less then either 
    0.5\,Gyr (upper panels) or 1.0\,Gyr (lower panels), as a 
    function of redshift. Left panels assume pure density 
    evolution in the quasar population above $z\sim2$, right 
    panels assume pure peak luminosity evolution.
    \label{fig:frac.young}}
\end{figure*}

Figure~\ref{fig:frac.young} considers the population of very recently
formed spheroids, which will not yet be relaxed or reddened and may be
identified as either peculiar or interacting galaxies.  We determine
the fraction of spheroids with ages less than 0.5\,Gyr (upper panels)
or 1.0\,Gyr (lower panels), as a function of redshift.  In the left
and right panels, we show this prediction assuming either pure density
or pure peak luminosity evolution of the quasar population above
$z\sim2$, respectively.  The predictions are similar for a limiting
age of both 0.5 and 1.0\,Gyr; that our predictions are not strongly
sensitive to the spheroid age in this regime suggests that this can be
observationally measured via relatively simple diagnostics.  Clearly,
direct measures of the population of merging and interacting galaxies
probe the fraction of galaxies with very recent formation times, but
by $\sim1$\,Gyr, many of these objects may be identified not through
more difficult morphological analysis but e.g.\ through spectral
classification as K+A galaxies.

Moreover, the fraction of young objects at e.g.\ $z=3$ is sensitive to
the strength of the density evolution modeled, which allows
observations of the distribution of spectral types as a function of
redshift to not only test our modeling but also to constrain the form
of high-redshift quasar evolution.  While at very low redshift the
results are similar, the prediction that the fraction of young objects
should be higher in low-mass spheroids reverses rapidly at $z\sim1$ in
the pure peak luminosity evolution case. This distinction should allow
even rough observations of the fraction of K+A vs.\ A galaxies at
$z\gtrsim1-2$ to break the observational degeneracy between pure
density and pure peak luminosity evolution.

\citet{BernardiIV} find from the color and chemical evolution of SDSS
elliptical galaxies that these galaxies are passively evolving at
redshifts $z\lesssim0.5$, and that they (on average) formed $\sim9$\
Gyr in the past. \citet{BernardiII,BernardiIII} determine the same
characteristic age independently based on analyses of the fundamental
plane and $z=0$ galaxy scaling relations. This corresponds to a
redshift of formation $z\sim1.5$, consistent with our predictions for
the formation redshifts of massive red galaxies.  This age also makes
it clear that the peak elliptical galaxy formation occurs
contemporaneously with peak quasar activity at $z\sim2$,
which is explained if
spheroids and quasars form together.

This is also consistent with direct observations of the morphologies
of galaxies, which show that by $z\sim0.7$ red galaxies are almost all
relaxing ellipticals, with little contribution to observed luminosity
from e.g.\ dusty spirals \citep{Bell04a}.  \citet{Fontana04} also find
similar results from studying ellipticals in the K20 survey; namely,
that massive ellipticals evolve passively for $z\lesssim0.7$, with
little growth in the total mass density in spheroids. However, at
$z\gtrsim1$, the mass growth in ellipticals rises steeply, with most
mass assembly at $z\sim1-2$.  Specifically, they estimate $\sim1/3$ of
the present mass of massive ellipticals has been assembled recently by
$z\sim2$, in agreement with our predictions for the evolution of the
stellar mass function and ages (Figures~\ref{fig:MF}, \ref{fig:ages},
and \ref{fig:frac.young}).  They further find that for $z\gtrsim1$,
the $z=0$ population of massive ellipticals becomes increasingly
dominated by star-forming galaxies, as expected in a merger-driven
scenario for contemporaneous spheroid and quasar formation.  Likewise,
\citet{Somerville04} and \citet{Daddi04} observe that at $z\sim1.5-2$,
the massive elliptical population includes large numbers of highly
disturbed morphologies indicative of merger-induced
starbursts. \citet{Cross04} find from fundamental plane analyses that
the production of massive red ellipticals should increase with cosmic
time to a peak at $z\sim2$ and then fall, suggesting that this is the
epoch of peak massive spheroid formation. This is also supported by
direct observations of quasar host galaxies, which find strong
evidence for simultaneous and strongly associated black hole growth
and star formation at redshifts corresponding to peak quasar activity
($z\gtrsim1$) \citep[e.g.,][]{Alexander05}.

Many observations indicate that galaxy age increases with velocity
dispersion or spheroid mass
\citep[e.g.,][]{Jorgensen99,Trager00,Kuntschner01,Caldwell03,
Fontana04,Bernardi05,Faber05,Howell05,Tanaka05,Gallazzi05,Nelan05}, as
we have considered in
Figure~\ref{fig:age.vs.sigma}. \citet{Gallazzi05} also quantify this
trend in terms of stellar mass, finding that galaxies with mass
$\sim10^{9}-10^{12}\,M_{\sun}$ form at redshifts $z\sim1.5-2$, with
median age increasing systematically with mass; they estimate e.g.\
$\sim16\%$ of $10^{12}\,M_{\sun}$ galaxies (at which point their
sample is spheroid-dominated) are in place by $z\sim2$, rising to
$\sim50\%$ at $z=1.8$, similar to our predictions in
Figure~\ref{fig:ages}.  This is a consequence of the strong
anti-hierarchical black hole growth implied by our interpretation of
the quasar luminosity function, where higher-mass black holes (thus
higher-$\sigma$ spheroids) form at higher redshift $z\sim2$, and thus
we reproduce both the mean age of $z=0$ spheroids and its evolution
with velocity dispersion and mass. These authors also find that higher
velocity dispersion does not imply strongly decreasing metallicity,
which is consistent with our picture of rapid metal enrichment (even
at high redshift) in the starburst phase of the merger.

Our results are consistent with ages inferred from fundamental plane
analyses
\citep[e.g.,][]{VF96,Jorgensen96,Jorgensen99b,vanDokkum98,vanDokkum99,vanDokkum00,vanDokkum01,
Treu01,Treu02,Gebhardt03,Cross04,Wuyts04,vandeVen03}, color and
spectral analyses \citep[e.g.,][]{Bower92,Ellis97,
Bernardi98,Stanford98,Ferreras99,Schade99,
Menanteau01,Kuntschner02,Treu02,Pozzetti03,vandeVen03,Bell04b,FS04,Labbe05},
and gravitationally lensed objects
\citep[e.g.,][]{Rusin03,Rusin05}. These all indicate typical formation
redshifts $z\sim1.5-2.5$, with a large range of formation redshifts 
$\Delta z\sim1.5-2.0$
\citep{Treu01,Treu02,vandeVen03,Cross04,Rusin05}, and subsequent
passive evolution of reddening remnant ellipticals.  Although
semi-analytical models of hierarchical galaxy formation reproduce this
as a general trend in the star formation history of the Universe,
recent results by \citet{Menci05}, which attempt to reproduce the
observed bimodal color distribution of galaxies, predict that red
galaxies formed only in dense environments, underpredicting the
relative red field galaxy population and the number of faint red
galaxies.  Furthermore this semi-analytical modeling predicts that red
galaxies form at much too high a redshift, $z\sim4-5$.  Explicitly,
\citet{Daddi04} find that the number density of massive spheroids
which are forming and should appear as highly disturbed starbursting
galaxies at $z\sim2$ is underpredicted by a factor of at least
$\sim30$ by current semi-analytical models.

A key ingredient in resolving this discrepancy is clear from the
results of \citet{SDH05a}, who show that feedback from black hole
growth and quasar activity is critical in rapidly terminating star
formation, allowing the production of quiescent red ellipticals even
from mergers of relatively low-mass (faint) objects at much lower
redshifts and explaining the observations of more recent formation
redshifts $z\sim2$.  Furthermore, the presence of a massive black hole
is also important in {\em maintaining} continued reddening of the
elliptical, as feedback from residual accretion can re-heat the gas,
suppressing further star formation after the merger. This is also
suggested directly by the comparison between luminosity functions and
the modeling of \citet{Nagamine01}, \citet{Menci04}, and
\citet{Granato04} in \citet{Fontana04}, who show that these models
under or over-predict the bright luminosity function at high redshift,
but that AGN feedback can regulate the slope of the galaxy stellar
mass function at low masses. It is also important to note that even
those models which incorporate black hole growth and feedback
\citep[e.g.][]{Granato04} must properly model the quasar lifetime and
its dependence on luminosity \citep{H05c,H05e} in order to
simultaneously reproduce the quasar and red galaxy luminosity
functions and other properties in any picture of merger-driven AGN
activity, as we demonstrate in Figures~\ref{fig:LF.B},
\ref{fig:color.mag.tracks}, and \ref{fig:ML.simplelife}.

\section{Conclusions}
\label{sec:conclusions}

Here, we have considered the consequences of a merger-driven scenario
for the joint formation of spheroids, quasars, and relic supermassive
black holes for the population of red galaxies.  As we demonstrate
elsewhere, the remnant spheroid hot X-ray emitting gas properties
\citep{Cox05}, morphologies (Cox et al. 2005b, in preparation),
metallicities (Cox et al. 2005c, in preparation), $\msigma$ relation
\citep{DSH05}, fine structure (e.g. Hernquist \& Spergel 1992), and
fundamental plane relations (Robertson et al.\ 2005c, in preparation)
agree with observations.  The expulsion of gas in these final stages
of black hole growth is violent, and leaves a gas-poor remnant, with
most of the remaining gas heated to virial X-ray emitting temperatures
and effectively terminating star formation.  This produces the
observed red, elliptical galaxy population in the bimodal
color/morphology distribution of galaxies, explaining the bimodality
seen at low and moderate redshifts with quasar feedback providing the
necessary means of quickly moving galaxies from the ``blue''
evolutionary sequency (with continual star formation) to the ``red''
sequence (with negligible ongoing star formation) \citep{SDH05a}.

We use our model of quasar lifetime and evolution in mergers derived
from simulations to de-convolve the observed quasar luminosity
function and determine the rate of formation of black holes of a given
final mass as a function of black hole mass and redshift.  Identifying
quasar activity with the formation of spheroids in the framework of
the merger hypothesis of hierarchical theories of galaxy formation, we
then determine the corresponding rate of formation of spheroids with
given properties as a function of redshift. 

We predict the distributions of galaxy velocity dispersions, the
galaxy mass function, mass density, and star formation rate, the
luminosity function in many observed wavebands (e.g., NUV, U, B, V, R,
r, I, J, H, K), the total number density and luminosity density of
galaxies, the distribution of colors as a function of magnitude for
several different wavebands, the distribution of colors as a function
of velocity dispersion, the distribution of mass to light ratios as a
function of mass, the luminosity-size relations, and the typical ages
and distribution of ages (formation redshifts) as a function of mass,
velocity dispersion, and luminosity. For each of these quantities, we
predict the evolution from redshift $z=0-6$, although at high
redshifts $z\gtrsim2$, our modeling suffers from the degeneracy
between pure peak luminosity evolution and pure density evolution in
the observed quasar luminosity function. Still, our results agree well
with observations over a wide range of redshifts.

Many of these predicted quantities, including the colors,
mass-to-light ratios, and luminosity-size relations of spheroids, are
essentially probes of the distribution of ages as a function of
spheroid mass. However, this does not mean that they are trivially
related, as they manifest a different dependence on subsequent star
formation, structural galaxy scalings (e.g.\ $\mbul-\mvir$ or
$\mbul-R_{e}$ relations), and dispersion in age as a function of
different variables (as for example we have shown that the dispersion
in colors and ages is different as a function of luminosity, mass, and
velocity dispersion).  Furthermore, if effects such as dry merging or
metallicity scaling with stellar mass were not, as we have
demonstrated, second order effects, they would break the implicit
self-consistency of these quantities.  Most important, different
samples which probe e.g.\ different mass ranges, environments, sample
sizes, and redshifts (and, correspondingly, have different systematic
effects and biases) measure different quantities and constrain age
distributions by these different methods, and therefore it is
important to compare to the complete range of such observations rather
than one particular choice.

Our results tie together the observed red, elliptical galaxy
population and the quasar and relic supermassive black hole
populations.  With our modeling of quasar and merger activity derived
from hydrodynamical simulations, we have shown that the diverse set of
{\em galaxy} observations listed above can be predicted directly from
the observed {\em quasar} luminosity function. We have demonstrated
that the quasar luminosity function implies the properties of the red
galaxy population and their evolution with redshift, providing
compelling evidence that spheroid and quasar formation must be driven
by the same process of galaxy merging.

Our methodology depends only on the form of the quasar lifetime as a
function of peak luminosity, and simple scaling relations between
black hole and galaxy properties such as the $M_{\rm BH}-\sigma$
relation.  Our simulations reproduce these scalings, independent of a
wide range of host galaxy properties including gas fractions, presence
or absence of bulges, initial black hole masses, ISM gas equation of
state, galaxy orbital parameters, and virial velocities.  For example,
we have varied the mass ratio of the merging galaxies and find that
these scalings are unchanged between simulations with mass ratios of
1:1, 2:1, 3:1, and 5:1. We demonstrate in \citet{H05f} that the
scaling of quasar lifetime with luminosity and peak luminosity can be
understood as a consequence of black hole self-regulation.  Thus, as
long as black holes still self-regulate in a manner which preserves
observed relations, we expect these scalings to be robust with respect
to mass ratios and the merger parameters listed above.

The independence of these scalings, expressed in this manner, has the
advantage that it allows us to relate and predict the properties of
the quasar, black hole, and spheroid populations independent of a
complete cosmological framework. Our approach thus allows us to
determine, without introducing tunable parameters or additional
uncertainty regarding detailed cosmological distributions, whether the
merger hypothesis and the joint formation of spheroids and
supermassive black holes in a quasar phase in major mergers are
simultaneously consistent with quasar and spheroid observations.
Furthermore, it allows us to constrain the underlying cosmological
rate of creation or formation of spheroids and quasars in major
mergers as a function of e.g.\ quasar peak luminosity or spheroid
mass. These constraints appear to be consistent with observational
estimates of merging galaxy luminosity functions (Hopkins et al.\
2005g, in preparation), as these have a well-defined peak and turnover
corresponding to that predicted in e.g.\ our $\nbul$ distribution
\citep[e.g.,][]{Xu04,Wolf05}.

Our detailed results for individual galaxy mergers and
constraints on the formation rates of spheroids can be combined with
and used to test cosmological models, but we caution against too
direct a comparison of predicted merger rates with the constraints
from our modeling, at least presently.  The mergers which produce
quasars and spheroids, and are therefore of interest to and
constrained by our modeling, are mergers not just of halos, but halos
that host galaxies, and where the galaxies themselves have comparable
masses and large reservoirs of cold gas, and will themselves merge in
a Hubble time. There are certainly sufficient halo-halo major mergers 
in the standard CDM cosmology 
to explain the galaxy merger rates we infer; for example, 
the calculations of e.g.\ \citet{KH00,WL03,Granato04} show 
that there are more than enough major mergers at all masses to 
account for observed quasars with a one-to-one
correspondence between quasars and ongoing halo mergers, even 
with a short quasar lifetime ${\rm d}t/{\rm d}\log L\sim 10^{7}\,$yr 
(much shorter than the quasar lifetime we calculate for luminosities 
below the break in the observed luminosity function).
However, cosmological simulations do not yet have the
resolution to determine the rates and properties of such mergers, let
alone the gas physics of star formation and black hole accretion and
feedback.  Semi-analytical models do not calculate the physics of
these processes in a self-consistent manner, and must adopt a number
of assumptions about merger properties which introduce considerable
uncertainty (and allow considerable fine-tuning) in the predictions of
the rates and effects of such mergers. 
Still, ideally, our results can
be combined with such approaches in a manner which greatly increases
their effective dynamic range, eventually enabling an a priori
prediction of the relevant merger rates and quasar and spheroid
properties from a fully theoretical framework.

The merger hypothesis presented by Toomre (1977) met with a great deal
of skepticism, much of which persists nearly 30 years later.  
However, many of the objections to Toomre's proposal owe to an
inappropriate comparison between the properties of interacting
galaxies seen {\it locally}, and those of large ellipticals which, in
our model, formed when the Universe was only a small fraction of its
present age.  For example, Ostriker (1980) argued that ellipticals
could not form in the manner suggested by Toomre because ellipticals
are more concentrated than disks of local spirals.  This viewpoint can
be expressed most neatly in terms of phase space densities:
ellipticals have higher central phase space densities than disks of
local spirals and because, according to Liouville's Theorem, phase
space density is conserved during a collisionless process, mergers
between disks cannot explain the high phase space density of
ellipticals (Carlberg 1986, Gunn 1987).  N-body simulations show that
this is indeed the case (e.g.\ Barnes 1988, 1992; Hernquist 1992, 1993a),
but this argument is flawed when applied to the merger hypothesis in
at least two ways.  First, disks at high redshifts were likely
more compact than their counterparts in the local Universe.  Second,
and perhaps more important, disks at $z>1$ were almost certainly
more gas-rich than those of local spirals.  As emphasized by e.g.\
Lake (1989), Liouville's Theorem does not apply to mergers involving
gas-rich galaxies because gas can radiate energy.

Previous efforts to include gas dissipation in galaxy mergers, such as
those of, e.g., Hernquist (1989), Barnes \& Hernquist (1991, 1996) and
Mihos \& Hernquist (1996) were restricted to cases where the
progenitor galaxies were $\sim 10\%$ gas because the ISM was modeled
as a single-phase, isothermal medium.  However, based on simulations and
simple physical arguments, Hernquist et al.\ (1993) estimated that
remnants of disk mergers would have a sufficiently high phase space
density to explain central properties of ellipticals only for
progenitor gas fractions $\gtrsim 25-30\%$.  More complex, more
realistic treatments of the ISM as a multiphase medium (e.g.  Springel
\& Hernquist 2003a) now make it possible to construct disks with much
larger gas fractions that do not violate the Toomre (1964) stability
criterion (see, e.g., Fig. 6 of Springel et al. 2005b).

The simulations used in the present study, which employ galaxies with
larger gas fractions than in earlier works and with galaxy structure
reflecting cosmic evolution, show that mergers can, in fact, account
for observed properties of ellipticals.  Furthermore, by incorporating
black hole growth and feedback into the simulations, we have
demonstrated that the various processes attending a gas-rich merger
can explain a much broader class of phenomena than Toomre's (1977)
original hypothesis.  Indeed, it is a remarkable fact that the
critical gas fraction suggested by Hernquist et al. (1993) to overcome
the phase space density problem is similar to that required for
mergers to produce AGN with luminosities matching those of bright
quasars at $z\sim 2$ as well as reproducing observed kinematic and
structural properties of ellipticals that have been puzzling up to
now.  For example, as we show in Cox et al. (2005b, in preparation),
the observed distribution of projected misalignments between spin and
minor axes of ellipticals is naturally reproduced by our models if the
gas fraction is large enough, which is not true for mergers between
gas-poor spirals. These gas fractions are appropriate for the 
redshifts of formation we have determined here, with most large
ellipticals building up their mass at moderate to high redshifts
$z\sim1.5-2.5$, and subsequent mergers primarily ``dry'' or collisionless. 
These various lines of evidence all support the
picture that quasars and ellipticals originated through the same
process; mergers between gas-rich galaxies.

Semi-analytical models in which interactions and galaxy mergers fuel
starburst activity \citep[e.g,][]{Cole00,Somerville01,Menci04} and
cosmological hydrodynamical simulations
\citep[e.g.,][]{Dave02,Nagamine04b,
Nagamine05a,Nagamine05b,Night05,Finlator05} have improved our
understanding of galaxy formation and evolution, reproducing the
properties of the cumulative galaxy population and explaining the
tendency of larger galaxies to be redder and older as a natural
consequence of hierarchical growth scenarios. Such modeling may even
be able to account for bimodality in the $z\lesssim1-2$ galaxy color
distribution \citep{Menci05}, with red galaxies formed in dense
environments at high redshifts $z\sim4-5$, with several early merging
events and interactions ceasing at later redshifts in these
environments.  However, as \citet{SDH05a} and this work make clear,
these models must incorporate feedback from AGN activity \citep[as in,
e.g.][]{Granato04} and the corresponding very rapid expulsion of gas
and quenching of star formation in mergers to explain the formation of
red spheroids at much later times $z\sim1.5-2$, as the bulk of
observations suggest (see \S~\ref{sec:ages}), as well as the
significant faint population of such objects and their field
population, as most observations find very little dependence on
environment in the red galaxy population at fixed mass or luminosity
\citep{Blanton03,Balogh04,Hogg04}.  Feedback from starburst-driven
winds and AGN may also be critical in suppressing excessive early
formation of low-mass spheroids \citep[e.g.][]{Granato04,Silva05}, in
order to explain the anti-hierarchical growth of spheroid and black
hole mass implied by the quasar luminosity function.  A proper
accounting of the luminosity dependence of the quasar lifetime shows
that the anti-hierarchical ``downsizing'' seen in both spheroid and
quasar evolution is completely self-consistent, which is {\em not} the
case if this dependence is ignored.

Also, unlike these and other previous galaxy evolution models, we are
able to specifically predict the properties of the red/spheroid
population, and do so without the addition of new tunable parameters. 
The input physics of our simulations and modeling is already strongly 
constrained by an extensive range of observations of 
quasar properties (Di Matteo et al.\ 2005; Robertson et al.\ 2005b; Hopkins et al.\ 2005a-e), 
essentially fixing our model, at which point the only essential observational input is the
observed {\em quasar} luminosity function.  Our predictions 
demonstrate that the observed properties of quasars provide powerful constraints 
on the spheroid population, and likewise that spheroid observations can strongly 
constrain quasar evolution, especially at low luminosity and high redshift where 
direct observations are difficult. We further demonstrate
that these predictions are skewed by several orders of magnitude if we
adopt idealized models of the quasar lifetime in which
quasars turn ``on''/``off'' or follow exponential light curves,
instead of the more complicated quasar evolution we have studied in
our simulations, demonstrating that it is not possible to reconcile
the quasar and spheroid galaxy luminosity functions or spheroid ages, colors, or 
mass-to-light ratios in models of joint
AGN-spheroid formation without accounting for
luminosity-dependent quasar lifetimes. As a result, previous attempts to infer the 
properties of the spheroid population from the quasar luminosity function 
\citep[e.g.,][]{MRD04}, although providing strong evidence of general co-evolution, 
have been forced to invoke evolution 
in e.g.\ the $\mbh-\mbul$ relation to explain integrated properties 
of the spheroid population and could not predict e.g.\ spheroid luminosity functions, 
whereas the application of more realistic quasar lifetimes 
immediately resolves these difficulties. Any modeling which attempts to
simultaneously reproduce the properties of quasars and galaxies
\citep[e.g.,][]{KH00,V03,WL03,Granato04}, specifically the remnant
spheroid population, with AGN activity triggered in interactions and
mergers must account for the effects of feedback and gas physics on
quasar evolution, and in particular must account for the non-trivial,
luminosity-dependent nature of the quasar lifetime.

We provide a large number of new predictions of the evolution of red
galaxy properties with redshift, for comparison with future
observations, which can be used to test this model and refine our
understanding of joint spheroid and AGN formation. Our modeling also
motivates observations of e.g.\ the ages, mass-to-light ratios, and colors of
low-mass/luminosity galaxies at $z=0-1$ to strongly constrain whether
pure luminosity or pure density evolution occurs in the
quasar/spheroid population at high redshift, where direct observations are
inaccessible. These observations can also constrain the shape of the
faint-end peak luminosity distribution, i.e.\ the low mass slope of
the rate at which quasars of a given final black hole mass form (directly
related to the remnant spheroid properties), where observations of
e.g.\ the quasar luminosity function provide only very weak
constraints \citep{H05e}.

\acknowledgments 

We thank an anonymous referee for comments that greatly improved the paper,
and Marijn Franx for generous assistance, suggestions, 
and discussion throughout the development of this paper. Our thanks also go to 
Marc Davis, Doug Finkbeiner, Sandy Faber, Jeff Newman, and Russell Smith, for 
helpful suggestions for the content and focus of this paper, and to Tiziana Di 
Matteo for discussion and development of the simulations upon which this 
work is based. 
This work was supported in part by NSF grants ACI 96-19019, AST 00-71019, AST
02-06299, and AST 03-07690, and NASA ATP grants NAG5-12140,
NAG5-13292, and NAG5-13381.  The simulations were performed at the
Center for Parallel Astrophysical Computing at the Harvard-Smithsonian
Center for Astrophysics.

\end{document}